\documentclass[11pt,a4paper]{article}

\usepackage[T1]{fontenc}
\usepackage[utf8]{inputenc}
\usepackage[a4paper,
    top=25mm,
    bottom=27mm,
    left=28mm,
    right=28mm,
    headheight=14pt,
    headsep=7mm,
    footskip=12mm
]{geometry}

\usepackage{silence}
\usepackage{jheppub}
\usepackage{microtype}
\usepackage{amsmath,amssymb}
\usepackage{lmodern}
\usepackage[dvipsnames]{xcolor}
\usepackage{graphicx}
\usepackage[
    compatibility=false,
    font=small,
    labelfont=bf,
    labelsep=period
]{caption}
\usepackage{subcaption}
\usepackage{titlesec}
\usepackage{tocloft}
\usepackage{fancyhdr}
\usepackage{float}
\usepackage{booktabs}
\usepackage{multirow}
\usepackage{siunitx}
\usepackage[table]{xcolor}
\usepackage{eso-pic}

\definecolor{PaperBlue}{HTML}{244F62}
\definecolor{PaperGray}{HTML}{626A70}
\definecolor{PaperRule}{HTML}{B7C4CA}
\definecolor{EmailBlue}{HTML}{2D6682}

\numberwithin{equation}{section}

\titleformat{\section}
  {\large\bfseries\color{PaperBlue}}
  {\thesection}{0.75em}{}
\titleformat{\subsection}
  {\normalsize\bfseries\color{PaperBlue}}
  {\thesubsection}{0.75em}{}
\titleformat{\subsubsection}
  {\normalsize\itshape\color{PaperGray}}
  {\thesubsubsection}{0.75em}{}
\titlespacing*{\section}{0pt}{2.6ex plus 0.8ex minus 0.3ex}{1.2ex}
\titlespacing*{\subsection}{0pt}{2.2ex plus 0.6ex minus 0.3ex}{0.8ex}
\titlespacing*{\subsubsection}{0pt}{1.8ex plus 0.5ex minus 0.2ex}{0.6ex}

\usepackage[nameinlink,capitalise]{cleveref}

\fancypagestyle{plain}{%
    \fancyhf{}
    \fancyfoot[C]{\footnotesize\thepage}

}

\fancypagestyle{firstpage}{%
    \fancyhf{}
    \fancyhead[R]{\small MITP-26-045}
    \fancyfoot[C]{\footnotesize\thepage}

}

\newcommand{\ii}{\mathrm{i}}
\newcommand{\ee}{\mathrm{e}}
\newcommand{\GeV}{\mathrm{GeV}}
\newcommand{\MeV}{\mathrm{MeV}}

\newcommand{\SLO}{S_{\mathrm{LO}}}
\newcommand{\SGradNLO}{S_{\mathrm{grad,NLO}}}
\newcommand{\SGradNLOcsix}{S_{\mathrm{grad,NLO}+c_6}}

\title{The devil in the transition: NLO nucleation and the particle physics behind the PTA signal}
\newcommand{\paperauthors}{Cristina Puchades-Ib\'a\~nez \qquad Pedro Schwaller}
\date{}

\makeatletter
\renewcommand{\maketitle}{%
    \begin{center}
        \vspace*{-1.2em}
        {\color{PaperBlue}\rule{\textwidth}{0.8pt}\par}
        \vspace{1.15em}
        {\LARGE\bfseries\color{PaperBlue}\@title\par}
        \vspace{1.25em}
        {\large\paperauthors\par}
        \vspace{0.75em}
        {\small
        PRISMA$^{++}$ Cluster of Excellence and Mainz Institute for Theoretical Physics,\\
        Johannes Gutenberg University Mainz, 55099 Mainz, Germany\par}
        \vspace{0.55em}
        {\small
        \href{mailto:crpuchad@uni-mainz.de}{crpuchad@uni-mainz.de}
        \qquad
        \href{mailto:pedro.schwaller@uni-mainz.de}{pedro.schwaller@uni-mainz.de}\par}
        \vspace{1.0em}
        {\color{PaperRule}\rule{0.72\textwidth}{0.5pt}\par}
    \end{center}
    \vspace{0.8em}
}
\makeatother

\renewenvironment{abstract}
  {\begin{center}
   \begin{minipage}{0.88\textwidth}
   \small
   \begin{center}\bfseries\color{PaperBlue}Abstract\end{center}
   \noindent\ignorespaces}
  {\par\end{minipage}\end{center}}

\begin{document}

\AddToHookNext{shipout/foreground}{%
    \put(
        \LenToUnit{\dimexpr\paperwidth-15mm+1in\relax},
        \LenToUnit{\dimexpr-12mm-1in\relax}
    ){%
        \makebox[0pt][r]{\small MITP-26-045}%
    }%
}

\maketitle
\thispagestyle{plain}

\begin{abstract}
We quantify how the treatment of a strongly supercooled phase transition affects the particle-physics interpretation of a nanohertz gravitational-wave signal. For the classically conformal Abelian Higgs model, we calculate the nucleation rate at NLO in the dimensionally reduced theory and test the gradient expansion against explicit fluctuation determinants. A calibrated correction reproduces the determinant exponent with a mean absolute relative difference of $0.8\%$ and allows this information to be included throughout the parameter scan. We propagate the corrected rate through percolation, reheating and gravitational-wave production, and confront the resulting spectra with the NANOGrav 15-year and IPTA DR2 data. The corrected NLO calculation shifts the preferred gauge coupling by $9\%$ and $13\%$, respectively, and the input scale by $18\%$ for NANOGrav. After the determinant calibration, the largest theoretical uncertainties arise from uncertainties in the gravitational-wave spectra and the relation between $R_\star$ and $\beta/H$. 
\end{abstract}

\setcounter{tocdepth}{1}
\tableofcontents
\clearpage

\section{Introduction}
\label{sec:introduction}

Several pulsar timing arrays have reported evidence for a gravitational-wave background at nanohertz frequencies \cite{NANOGrav:2023gor,EPTA:2023fyk,Reardon:2023gzh,Xu:2023wog}.
Strongly supercooled dark-sector phase transitions provide a possible cosmological origin of this signal \cite{NANOGrav:2023hvm,Madge:2023dxc,Bringmann:2023opz}.
Explicit conformal realizations have also been shown to reproduce the PTA data \cite{Fujikura:2023lkn,Salvio:2023blb,Goncalves:2025uwh,Balan:2025uke,Bringmann:2026xcx}.
The question is therefore no longer whether such a transition can reproduce the PTA signal, but how reliably the underlying particle-physics parameters can be inferred from it.

This is the gravitational-wave inverse problem \cite{Madge:2023dxc}, and its accuracy is not determined by the data alone.
Connecting a Lagrangian to a PTA spectrum requires the nucleation rate, the percolation and reheating history, the characteristic bubble separation, the wall dynamics and the gravitational-wave source, which are not known to the same accuracy \cite{Croon:2020cgk}.
Systematic NLO treatments of thermal nucleation are now available \cite{Hirvonen:2021zej,Ekstedt:2021kyx,Kierkla:2025qyz,Christiansen:2025xhv}, while PTA analyses of explicit models generally rely on leading-order or semi-analytic bounce actions \cite{Goncalves:2025uwh,Bringmann:2026xcx}. 
The impact of the nucleation calculation on parameter reconstruction has recently been studied for LISA \cite{Kierkla:2026bnm}, but not yet in an inference from PTA data.
Carrying this accuracy through the transition history and into the PTA likelihood is therefore necessary to identify which theoretical uncertainties limit the reconstruction.

The classically conformal Abelian Higgs model provides a minimal setting in which to follow the theoretical uncertainty from nucleation to PTA inference.
In the scale-invariant limit, the transition is controlled by the gauge coupling $g$ and a single input scale $\mu_0$, while the radiatively generated barrier produces the strong supercooling relevant for PTAs \cite{Levi:2022bzt,Kierkla:2023von,Christiansen:2025xhv}.
We calculate the nucleation rate at NLO in the dimensionally reduced three-dimensional EFT and use explicit functional determinants to test the gradient expansion \cite{Hirvonen:2021zej,Ekstedt:2021kyx,Kierkla:2025qyz}.
Since the determinant calculation is too expensive for the full scan, the benchmark results are used to calibrate a correction that incorporates this information across the $(g,\mu_0)$ plane. The corrected rate is then propagated through percolation and reheating, including the temperature dependence of the relativistic degrees of freedom \cite{Saikawa:2018rcs,Li:2025nja}. We determine the characteristic bubble separation directly from the integrated bubble number density, rather than from the action derivative \cite{Turner:1992tz,Megevand:2016lpr,Matuszak:2026xsz}.
We include the wall dynamics following Ref.~\cite{Lewicki:2022pdb} and use the cosmic-expansion calculation of Ref.~\cite{Lewicki:2025hxg} as a benchmark for the uncertainty in the gravitational-wave source. Finally, we confront the predicted spectra with the NANOGrav 15-year and IPTA DR2 free-spectrum results using \textsc{Ceffyl} \cite{NANOGrav:2023hvm,Antoniadis:2022pcn,Lamb:2023jls}.

The paper is organised as follows. \Cref{sec:model} introduces the near-conformal dark sector and the role of the input scale $\mu_0$. \Cref{sec:nucleation-rate} develops the dimensionally reduced NLO nucleation rate, its calibration to the functional determinants, and the treatment of percolation and completion. The scan of transition parameters, the wall dynamics, and the gravitational-wave predictions, including the cosmic-expansion benchmark, are presented in \cref{sec:GW-spectrum}. \Cref{sec:phenomenology} compares these predictions with PTA data and studies how the theoretical treatment shifts the preferred model parameters. We conclude in \cref{sec:conclusions}. The appendices contain the model and dimensionally reduced potential (\cref{app:model-equations}), the comparison between transition prescriptions (\cref{app:comparison-plots}), and details of the numerical scan and PTA likelihood (\cref{app:Scan-GWanalysis}). The data files are available upon request, and will be made publicly available soon.

\section{Near-conformal dark-sector symmetry breaking}
\label{sec:model}

The Abelian Higgs model provides a minimal realisation of a dark-sector first-order phase transition and its associated gravitational-wave signal. The theory consists of a dark $U(1)_D$ gauge field $A_\mu$ and a single complex scalar
$\Phi=(\phi+\ii\chi)/\sqrt{2}$ with unit charge, and is described by

\begin{equation}
\mathcal{L}_{D}
=
-\frac{1}{4}F_{\mu\nu}F^{\mu\nu}
+\left|(\partial_\mu-\ii g A_\mu)\Phi\right|^2
+m^2|\Phi|^2
-\lambda|\Phi|^4 .
\label{eq:lagrangian}
\end{equation}
We set all renormalisable interactions with the Standard Model to zero, in particular the gauge-kinetic mixing and the Higgs-portal coupling.
The phase-transition dynamics are therefore determined entirely by the dark sector described in \cref{eq:lagrangian}.
As in \cite{Madge:2023dxc}, we assume that the released vacuum energy can subsequently be transferred to the visible sector through interactions that are sufficiently weak not to affect the phase-transition calculation. 
For the cosmological evolution we adopt the benchmark temperature ratio $\xi\equiv T_D/T_{\rm SM}=1$ for simplicity. 

The model is made classically scale-invariant by setting $m^2=0$, which provides a simple setup for strongly supercooled phase transitions \cite{Jinno:2016knw,Levi:2022bzt}. In the absence of an explicit mass scale the symmetry-breaking scale is generated dynamically. At weak coupling, this occurs through the Coleman--Weinberg mechanism \cite{Coleman:1973jx}. The radiatively generated potential evolves slowly with temperature, allowing nucleation to be delayed far below the critical temperature\footnote{See also \cite{Salvio:2026bco} for a recent review.}. The resulting large vacuum-energy release can source a strong gravitational-wave signal, which at sufficiently low transition scales may fall within the PTA frequency band \cite{Baldes:2018emh,Ratzinger:2020koh,Lewicki:2020jiv,Levi:2022bzt,Sagunski:2023ynd,Gouttenoire:2023pxh,Madge:2023dxc,Fujikura:2023lkn,Salvio:2023blb,Ferrante:2023bcz,Bringmann:2023opz,Balan:2025uke,Li:2025nja,Goncalves:2025uwh}. 
To be specific, we define the theory at an input scale $\mu_0$ where we set~\cite{Christiansen:2025xhv}
\begin{equation}
m^2(\mu_0)=0,
\qquad
\lambda(\mu_0)=0,
\qquad
g(\mu_0)\equiv g .
\label{eq:boundary-conditions}
\end{equation}
The running of the mass parameter is negligible, while the quartic coupling runs significantly and typically changes sign close to the scale of radiative symmetry breaking. Setting $\lambda(\mu_0)=0$ fixes $\mu_0$ as the scale at which the renormalised quartic vanishes, leaving $g$ and $\mu_0$ as the two independent input parameters of the dark sector. The relation between $\mu_0$ and the physical symmetry-breaking scale is made explicit below. It should be stressed that this is not a limitation of the model parameter space: $\lambda(\mu)$ always changes sign since the one-loop beta function is positive -- instead choosing $\lambda$ and $\mu_0$ independently would be an overcounting of parameters.

\subsection{Coleman--Weinberg scale and the role of $\mu_0$}
\label{subsec:CW-scale}

The zero-temperature one-loop Coleman--Weinberg potential has the general form~\cite{Coleman:1973jx}
\begin{equation}
V_{\rm CW}^{(1)}(\phi)
=
\sum_i
\frac{n_i\,m_i^4(\phi)}{64\pi^2}
\left[
\log\left(\frac{m_i^2(\phi)}{\mu^2}\right)
-c_i
\right],
\label{eq:CW-general}
\end{equation}
where $m_i(\phi)$ are the field-dependent masses, $n_i$ counts the corresponding
degrees of freedom, including the fermionic sign, and $c_i$ depends on the
renormalisation scheme.

For our conformal Abelian Higgs model, in the limit written in \cref{eq:boundary-conditions} and retaining the leading gauge contribution gives
\begin{equation}
V_{\rm CW}(\phi;g,\mu_0)
=
\frac{3g^4\phi^4}{64\pi^2}
\left[
\log\left(\frac{g^2\phi^2}{\mu_0^2}\right)
-\frac{5}{6}
\right].
\label{eq:CW-potential}
\end{equation}
The minimum of \cref{eq:CW-potential} is located at
\begin{equation}
\langle \phi \rangle \equiv v=
\frac{\ee^{1/6}\mu_0}{g},
\label{eq:CW-vev}
\end{equation}
and the corresponding vacuum-energy difference is
\begin{equation}
\Delta V_{\rm CW}
\equiv
V_{\rm CW}(0)-V_{\rm CW}(v)
=
\frac{3\ee^{2/3}}{128\pi^2}\mu_0^4 .
\label{eq:CW-deltaV}
\end{equation}
The explicit dependence on $g$ therefore cancels from $\Delta V_{\rm CW}$,
although the position of the minimum retains its $g$ dependence through
\cref{eq:CW-vev}. The two input parameters play distinct roles: $g$ controls the shape of the radiatively generated potential, while $\mu_0$ fixes its overall scale. At fixed $g$, the dark-sector potential can consequently be expressed in terms of $\phi/\mu_0$ and $T/\mu_0$, allowing the same dimensionless thermal calculation to be reused when scanning over $\mu_0$. The dependence introduced by the cosmological expansion rate will be included explicitly below.

For the finite-temperature calculation, the couplings are evolved away from $\mu_0$ using the one-loop renormalisation-group equations. Including this running is particularly important in near-conformal models, where the hierarchy of scales can otherwise induce a sizeable residual dependence on the renormalisation scale \cite{Croon:2020cgk, Schicho:2022wty, Kierkla:2023von}. The quartic illustrates the relevant behaviour,
\begin{equation}
\beta_\lambda
\equiv
\frac{\mathrm{d}\lambda}{\mathrm{d}\log\mu}
=
\frac{10\lambda^2-6g^2\lambda+3g^4}{8\pi^2}\,.
\label{eq:lamb-run}
\end{equation}
The quartic runs negative towards scales below $\mu_0$, playing an important role in determining the location and shape of the potential barrier \cite{Christiansen:2025xhv}. The complete set of renormalisation-group equations used in the numerical analysis is collected in \cref{app:model-equations}.

At finite temperature, gauge interactions generate a positive thermal mass
that stabilises the symmetric phase at the origin. At the same time, the
radiatively generated negative quartic favours the broken-field direction.
Together with the bosonic thermal contributions, this produces the barrier
separating the two phases and determines the tunnelling dynamics. In the
following, we construct the finite-temperature effective theory and determine
the corresponding nucleation dynamics, including the higher-order corrections
relevant for a controlled description of the transition.

\section{Thermal transition beyond leading order}
\label{sec:nucleation-rate}

\subsection{Dimensional reduction and perturbative organisation}
\label{subsec:DR}

The thermal transition involves several parametrically separated scales. While non-zero Matsubara modes carry momenta of order $\pi T$, the bosonic zero modes that control the transition are considerably softer. Dimensional reduction \cite{Laine_2016,Ekstedt:2022bff} makes this separation explicit: the hard modes are integrated out and their effects are encoded in the parameters of a three-dimensional EFT for the long-distance fields. Besides resumming the thermal masses, this construction provides a systematic power counting for the nucleation calculation \cite{Gould:2021ccf,Lofgren:2021ogg, Hirvonen:2021zej}. For the Abelian model considered here, the resulting EFT contains the complex scalar zero mode $\Phi_3$, containing the radial mode $\phi_3$ describing the phase transition and the Goldstone fluctuation, the spatial gauge field $A_i$, and the temporal gauge mode $A_0$.

Starting from the near-conformal input choice in \cref{eq:boundary-conditions}, we evolve the four-dimensional parameters to the thermal scale and perform the matching with \texttt{DRalgo} \cite{Ekstedt:2022bff}. Our default scale prescription is
\begin{equation}
    \mu_R=2\pi T,
    \qquad
    \mu_3=g(\mu_R)T,
    \label{eq:matching-scales}
\end{equation}
where $\mu_R$ is the four-dimensional matching scale and $\mu_3$ is the renormalisation scale of the three-dimensional theory. We perform the hard-scale matching at next-to leading order (NLO) using \texttt{DRalgo}. This includes two-loop thermal corrections to the scalar and Debye masses and one-loop corrections to the three-dimensional couplings.
Within the three-dimensional EFT, we adopt the radiative-barrier power counting \cite{Kierkla:2023von,Arnold:1992rz}
\begin{equation}
m_3^2\phi_3^2
\sim
\lambda_3\phi_3^4
\sim
\frac{\bigl(g_3^2\phi_3^2\bigr)^{3/2}}{\pi},
\label{eq:radiative-barrier-counting}
\end{equation}
under which the gauge-induced one-loop barrier contributes already at leading order. With this power counting, the one-loop three-dimensional potential determines the leading order (LO) bounce, while the two-loop contribution enters perturbatively at NLO:
\begin{equation}
    \widehat V_3^{\mathrm{1L}}(\phi_3;T)
    \equiv
    V_3^{\mathrm{1L}}(\phi_3;T)
    -V_3^{\mathrm{1L}}(0;T),
    \qquad
    \widehat V_3^{\mathrm{2L}}(\phi_3;T)
    =
    \widehat V_3^{\mathrm{1L}}(\phi_3;T)
    +\Delta V_{3,\mathrm{2L}}(\phi_3;T).
    \label{eq:potential-prescription}
\end{equation}
The one-loop potential retains the contributions from the spatial and temporal gauge modes and the scalar modes, although the barrier is dominated by the spatial gauge fields. The explicit potentials are collected in \cref{app:model-equations}.

\subsection{Thermal bubble nucleation beyond leading order}
\label{subsec:nucleation-prescriptions}

A first-order transition proceeds through the nucleation of bubbles of the
broken phase across the potential barrier separating it from the metastable
symmetric phase. Critical bubbles must nucleate at a sufficient rate to
compete with the expansion of the Universe. The relevant quantity is therefore
the nucleation rate per unit volume,
\begin{equation}
\Gamma(T)=A(T)\exp[-S(T)]\,,
\label{eq:thermal-nucleation-rate}
\end{equation}
where $S(T)$ \footnote{In the three-dimensional normalisation used below, $S(T)$ is dimensionless and corresponds to the conventional quantity $S_3(T)/T$.} is the tunnelling exponent and $A(T)$ contains the fluctuation and dynamical prefactors. 
We define the nucleation temperature through
\begin{equation}
\Gamma(T_n)=H^4(T_n)\,,
\label{eq:nucleation-condition}
\end{equation}
using the full Hubble rate, including both radiation and vacuum contributions \cite{Megevand:2016lpr,Ellis:2018mja},

\begin{equation}
H^2(T)=
\frac{\rho_{\mathrm{rad}}(T)+\Delta V_{\mathrm{CW}}}
{3M_{\mathrm{Pl}}^2}\,.
\label{eq:full-Hubble}
\end{equation}
where $\rho_{\rm rad}(T)=\pi^2g_{*\rho}(T)T^4/30$
with $g_{*\rho}(T)$ denoting the effective relativistic degrees of freedom
contributing to the radiation energy density.
Since the nucleation rate depends exponentially on the action, even formally subleading corrections can appreciably shift the transition temperature. This is particularly important for radiatively generated transitions, where the gauge fluctuations responsible for the barrier also enter the higher-order corrections to the tunnelling exponent \cite{Ekstedt:2022ceo}.

Let $\phi_b^{\rm LO}(r;T)$ denote the $O(3)$-symmetric critical bubble obtained from the normalised one-loop three-dimensional potential $\widehat V_3^{\mathrm{1L}}$, where $r$ is the distance from the center of the bubble. Its action is
\begin{equation}
\SLO(T)
=
4\pi\int_0^\infty \mathrm{d}r\,r^2
\left[
\frac{1}{2}
\left(
\frac{\mathrm{d}\phi_b^{\mathrm{LO}}}{\mathrm{d}r}
\right)^2
+
\widehat V_3^{\mathrm{1L}}
\bigl(\phi_b^{\mathrm{LO}};T\bigr)
\right].
\label{eq:SLO}
\end{equation}
At NLO, the two-loop potential and the leading corrections to the kinetic term are evaluated perturbatively on this same bounce, in the gradient expansion approximation \cite{Hirvonen:2021zej, Kierkla:2025qyz}
\begin{align}
\SGradNLO(T)
&=
\SLO(T)
+4\pi\int_0^\infty \mathrm{d}r\, r^2
\left[
\Delta V_{3,\mathrm{2L}}
\bigl(\phi_b^{\mathrm{LO}};T\bigr)
+
\frac{1}{2}
\left(Z_{\mathrm{sp}}+Z_0\right)
\left(
\frac{\mathrm{d}\phi_b^{\mathrm{LO}}}{\mathrm{d}r}
\right)^2
\right],
\label{eq:SGradNLO}
\\
\SGradNLOcsix(T)
&=
\SGradNLO(T)
+4\pi\int_0^\infty \mathrm{d}r\, r^2\,
\Delta V_{3,6}\bigl(\phi_b^{\mathrm{LO}};T\bigr).
\label{eq:SGradNLOc6}
\end{align}
Here $Z_{\mathrm{sp}}$ and $Z_0$ denote the gradient corrections generated by the spatial and temporal gauge modes, respectively, while $\Delta V_{3,6}$ contains the leading dimension-six operators. Refs.~\cite{Chala:2024xll,Bernardo:2025vkz} suggest that higher dimensional operators can become unsuppressed for strong phase transitions. We use $\SGradNLOcsix$ to test their impact in our classically conformal, supercooled regime and find the correction to be negligible over the parameter range considered, as quantified below.
We therefore do not include it in our default NLO prescription.
A more comprehensive analysis will be presented in~\cite{BKLSS}.
Explicit expressions and conventions are collected in \cref{app:model-equations}. Since the LO bounce extremises $\SLO$, the correction to the bubble profile itself contributes only at higher order. Evaluating the NLO terms on $\phi_b^{\rm LO}$ therefore gives the consistent order-by-order expansion of the action. For these prescriptions, we approximate the prefactor dimensionally by $A(T) \approx T^4$.

The gradient-expanded result provides a local description of how the fluctuations respond to the bubble background. Its validity requires the modes being integrated out to remain sufficiently heavy compared with the characteristic momentum scale of the bubble.
As emphasised in \cite{Ekstedt:2021kyx,Kierkla:2025qyz}, this hierarchy
is not maintained along the bounce profile. The spatial gauge mode is heavy in the
bubble interior but becomes light towards the symmetric-phase tail, where
$M_A(\phi_3)\to0$. Although the NLO gradient contribution to the action
remains finite, the behaviour $Z_{\mathrm{sp}}\propto1/|\phi_3|$ signals that
the derivative expansion is not uniformly controlled in this region. By
contrast, the temporal gauge mode remains Debye screened and does not suffer
from the same breakdown.

Alongside the gradient-expanded prescription, we also consider a
functional-determinant calculation, in which the one-loop fluctuations are evaluated around the LO bounce without a local derivative expansion \cite{Ekstedt:2021kyx,Kierkla:2025qyz},
\begin{equation}
    \Gamma_{\mathrm{NLO,det}}(T)
    =
    A_{\mathrm{dyn}}(T)\,
    \det\nolimits_{\mathrm{s}}
    \bigl[\phi_b^{\mathrm{LO}}\bigr]\,
    \det\nolimits_{\mathrm{V}}
    \bigl[\phi_b^{\mathrm{LO}}\bigr]\,
    e^{-\SLO(T)}\,
    e^{-\Delta S_{V,\mathrm{NLO}}(T)}\,
    e^{\mathcal C_{\mathrm{sub}}
    [\phi_b^{\mathrm{LO}}]} \,.
    \label{eq:determinant-rate}
\end{equation}
Here, $\det_{\mathrm{s}}$ denotes the scalar contribution, while
$\det_{\mathrm{V}}$ contains the temporal, transverse and mixed
gauge--Goldstone sectors and $\mathcal C_{\mathrm{sub}}$ subtracts the corresponding zero-momentum contribution already included in $\SLO$. The zero-momentum NLO correction is the two-loop term already appearing in \cref{eq:SGradNLO},
\begin{equation}
    \Delta S_{V,\mathrm{NLO}}(T)
    \equiv
    4\pi\int_0^\infty \mathrm{d}r\,r^2\,
    \Delta V_{3,\mathrm{2L}}
    \bigl(\phi_b^{\mathrm{LO}};T\bigr)\,.
    \label{eq:DeltaSVNLO}
\end{equation}
\begin{figure}[t]
    \centering
    \begin{minipage}{0.48\textwidth}
        \centering
        \includegraphics[width=\textwidth]{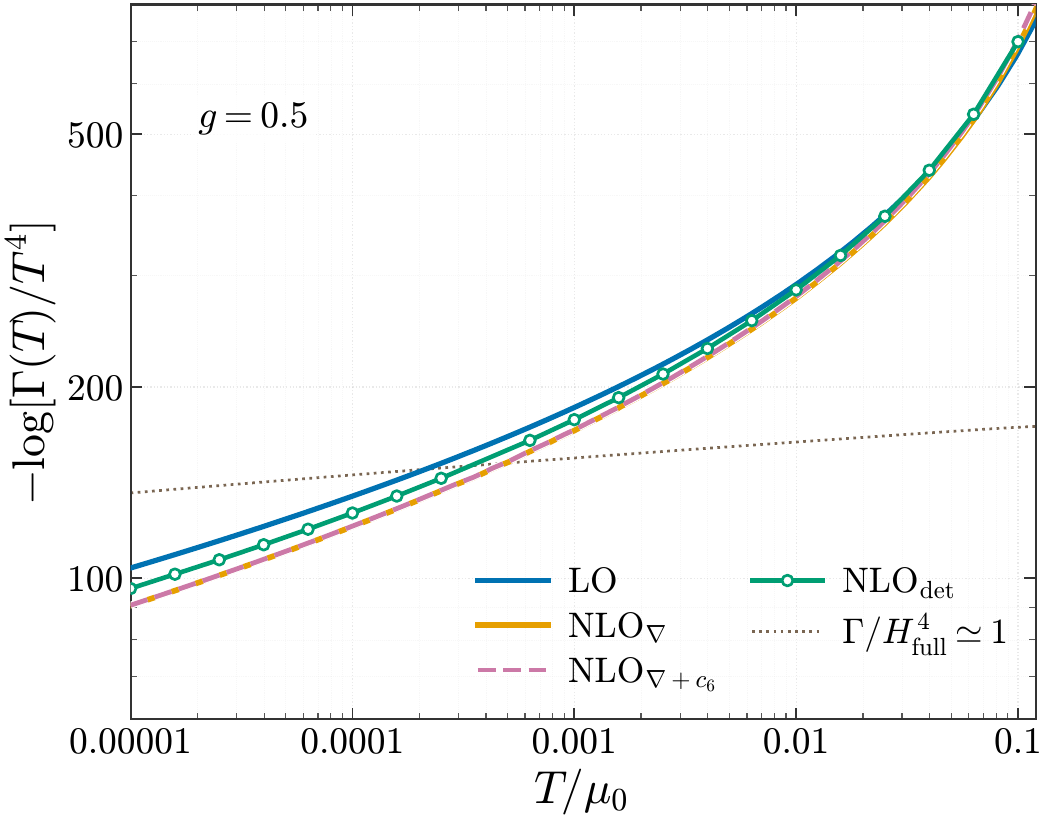}
        \par\vspace{1mm}
        {\small (a) $g=0.5$: action.}
    \end{minipage}
    \hfill
    \begin{minipage}{0.48\textwidth}
        \centering
        \includegraphics[width=\textwidth]{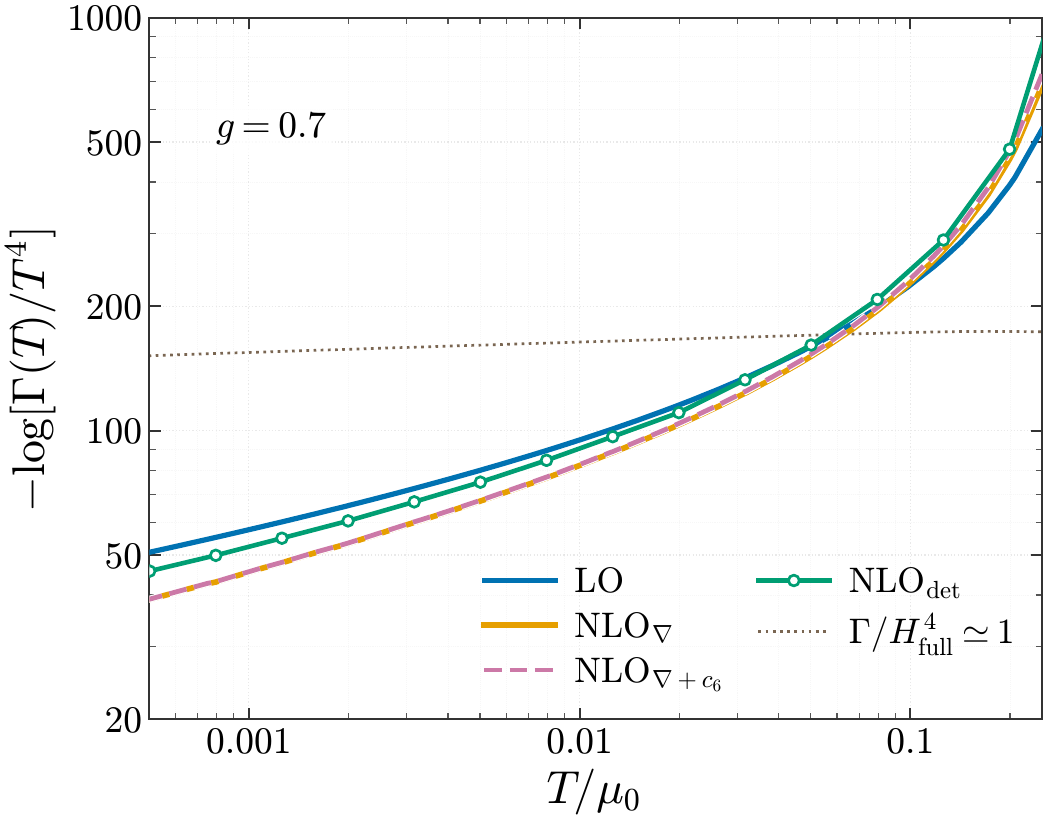}
        \par\vspace{1mm}
        {\small (b) $g=0.7$: action.}
    \end{minipage}

    \vspace{0.25cm}

    \begin{minipage}{0.48\textwidth}
        \centering
        \includegraphics[width=\textwidth]{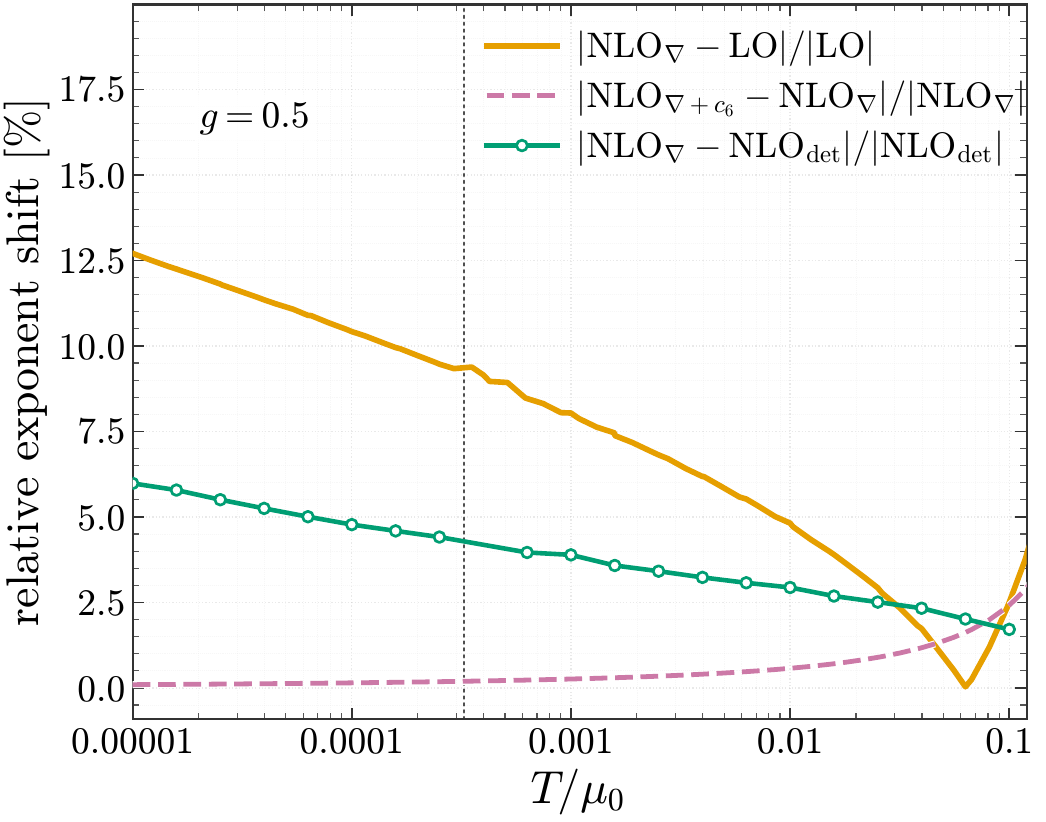}
        \par\vspace{1mm}
        {\small (c) $g=0.5$: relative corrections.}
    \end{minipage}
    \hfill
    \begin{minipage}{0.48\textwidth}
        \centering
        \includegraphics[width=\textwidth]{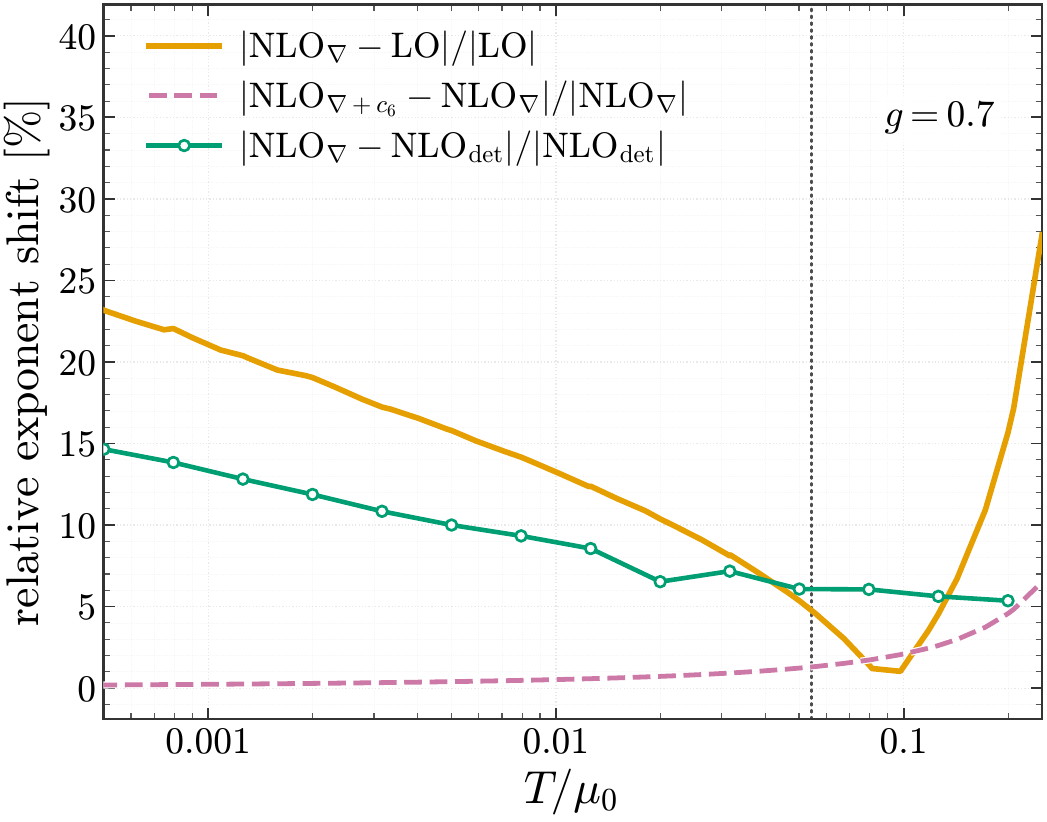}
        \par\vspace{1mm}
        {\small (d) $g=0.7$: relative corrections.}
    \end{minipage}

    \caption{
Comparison of the nucleation-rate exponent $-\log[\Gamma(T)/T^4]$ for $g=0.5$ (left) and $g=0.7$ (right). The upper panels show the LO result (blue), the NLO gradient expansion without and with the $c_6$ contribution (orange and dashed magenta), and the full determinant result (green circles). The dotted brown curve marks $\Gamma/H_{\rm full}^4=1$. The lower panels show the corresponding relative shifts: NLO versus LO (orange), the $c_6$ correction (magenta), and the determinant correction relative to GradNLO (green). The vertical dotted line indicates the GradNLO percolation temperature $T_p^{\nabla}$.
}
\label{fig:rate-determinant-comparison}

    \label{fig:action-corrections}
\end{figure}

We use this comparison to assess the accuracy of the gradient prescription and how differences with the determinant result affect the gravitational-wave spectrum.
Since the computation of the full fluctuation determinants is computationally expensive, we use the available benchmark points to construct a correction to the gradient result.

\Cref{fig:action-corrections} compares the different prescriptions for the nucleation rate at $g=0.5$ and $g=0.7$. Around the percolation temperature, the NLO potential and gradient corrections lower the bounce action relative to LO and therefore enhance the nucleation rate. The correction is larger for the smaller coupling: at $T_p$, the exponent changes by approximately $9.2\%$ for $g=0.5$ and $4.6\%$ for $g=0.7$.

In the conformal limit \cite{Bernardo:2026whs} found sizeable effects in equilibrium observables at $T_c$, while noting that these effects should be weaker for observables evaluated at the nucleation or percolation scale. 
In \cite{Christiansen:2025xhv}, we found no visible effect on the potential barrier around $T_n$, but did not evaluate the correction to the tunnelling exponent or percolation.
This distinction is realised explicitly here: nucleation and percolation occur far below $T_c$, and the bounce explores field values well below the broken-phase minimum, suppressing the dimension-six contribution. Accordingly, the leading dimension-six operator changes the nucleation exponent by only $0.2\%$ for $g=0.5$ and $1.3\%$ for $g=0.7$. It is therefore negligible in the supercooled region relevant for our analysis, and we omit it from the remaining calculation.\footnote{While $\Delta V_{3,6}$ is the only dimension 6 operator that enters the effective potential directly, other operators at this order enter via the fluctuation determinant. We expect their impact to be of similar order. A comprehensive study will be presented in~\cite{BKLSS}.}

We finally use the functional determinants to test the local gradient treatment of the NLO fluctuations. This is not an additional perturbative order, but a benchmark of the gradient approximation. Around $T_p$, replacing the gradient-expanded result by the functional determinants changes the nucleation exponent by approximately $4.3\%$ for $g=0.5$ and $6.1\%$ for $g=0.7$. The gradient prescription gives the smaller exponent and therefore predicts earlier percolation than the determinant calculation. 

In the strongly supercooled regime, the LO result can lie closer to the functional determinant than the NLO gradient-expanded result.
The difference is mainly driven by the spatial gauge--Goldstone contribution.
We therefore use the determinant results to define a simple correction to the spatial gradient contribution.
We write the contribution of the spatial gauge boson in \cref{eq:SGradNLO} as $Z_{\rm sp}$, and we use
\begin{equation}
    S_{\rm corr}(T)
    =
    S_{\rm grad,NLO}(T)
    -
    K\,g^p
    \left(\frac{T}{\mu_0}\right)^q
    S_{Z_{\rm sp}}(T).
    \label{eq:corrected-gradient-action}
\end{equation}

A fit to the determinant benchmarks gives $K=0.617$, $p=1.369$ and $q=-0.0351$.
Details of the procedure are given in \cref{app:gradient-correction}.
We use \cref{eq:corrected-gradient-action} as our central nucleation prescription for the results of this work.

\begin{figure}[t]
    \centering

    \begin{subfigure}[t]{0.48\textwidth}
        \centering
        \includegraphics[width=\linewidth]{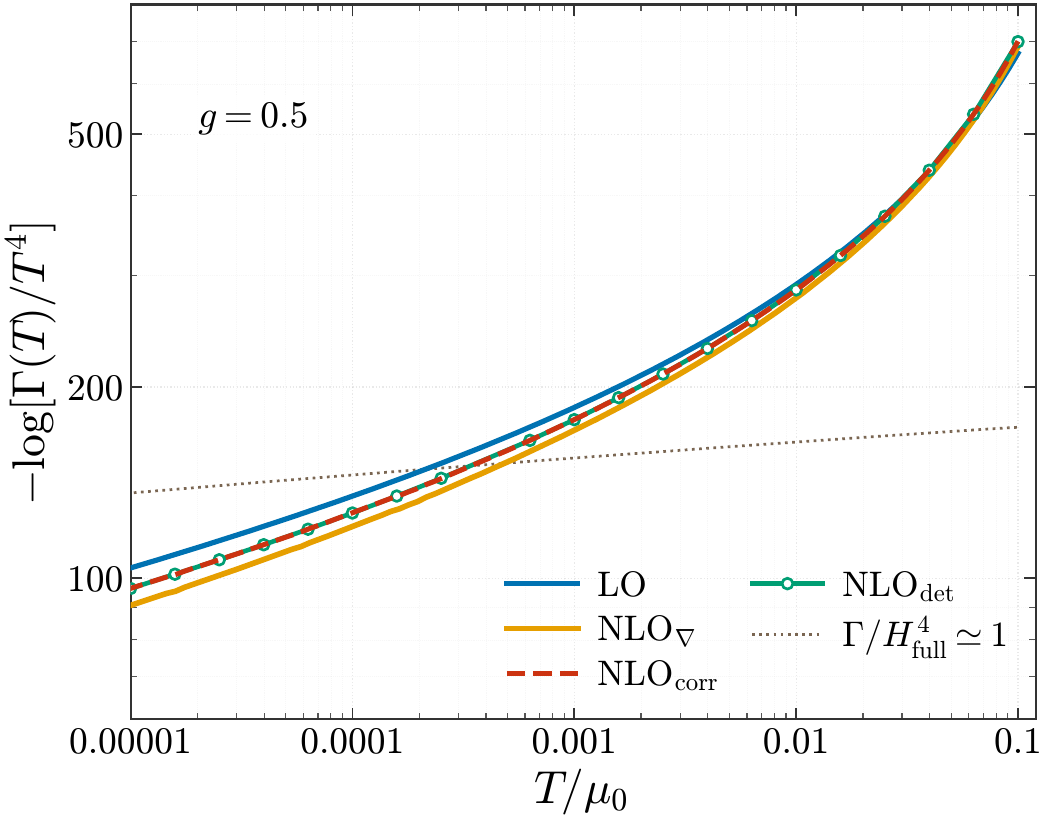}
        \caption{$g=0.5$: nucleation-rate exponent.}
        \label{fig:actioncorr-correction-g05}
    \end{subfigure}
    \hfill
    \begin{subfigure}[t]{0.48\textwidth}
        \centering
        \includegraphics[width=\linewidth]{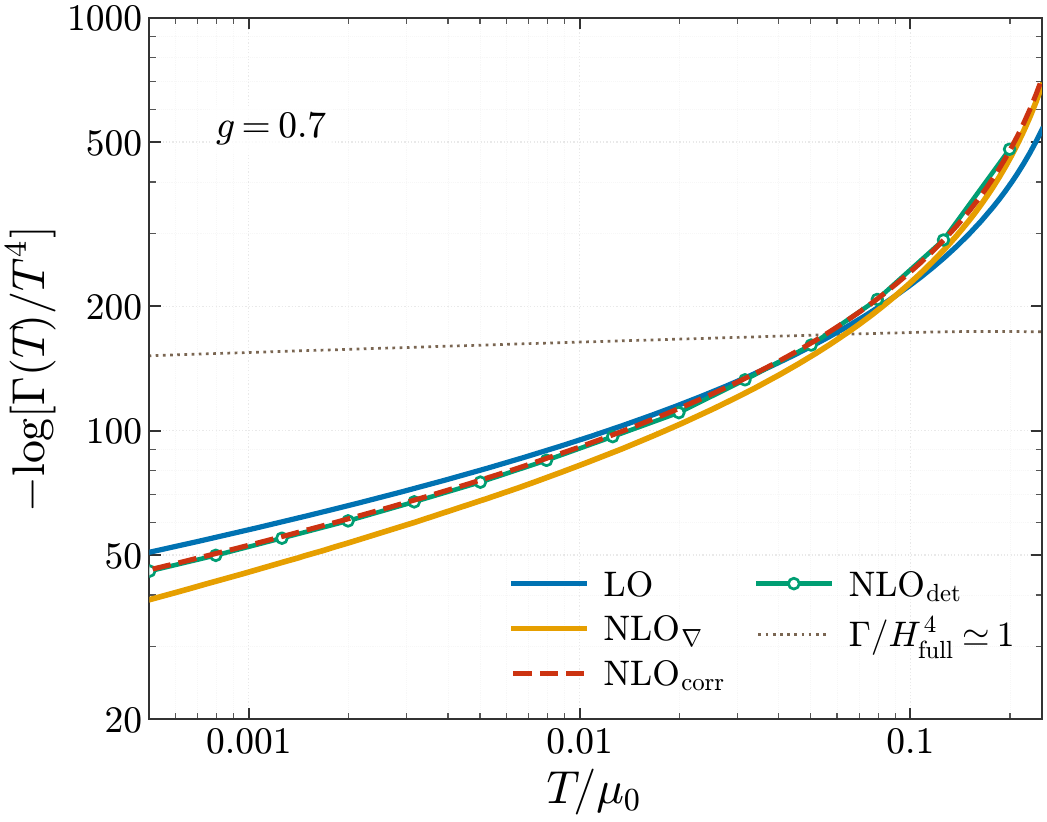}
        \caption{$g=0.7$: nucleation-rate exponent.}
        \label{fig:actioncorr-correction-g07}
    \end{subfigure}

    \vspace{0.5em}

    \begin{subfigure}[t]{0.48\textwidth}
        \centering
        \includegraphics[width=\linewidth]{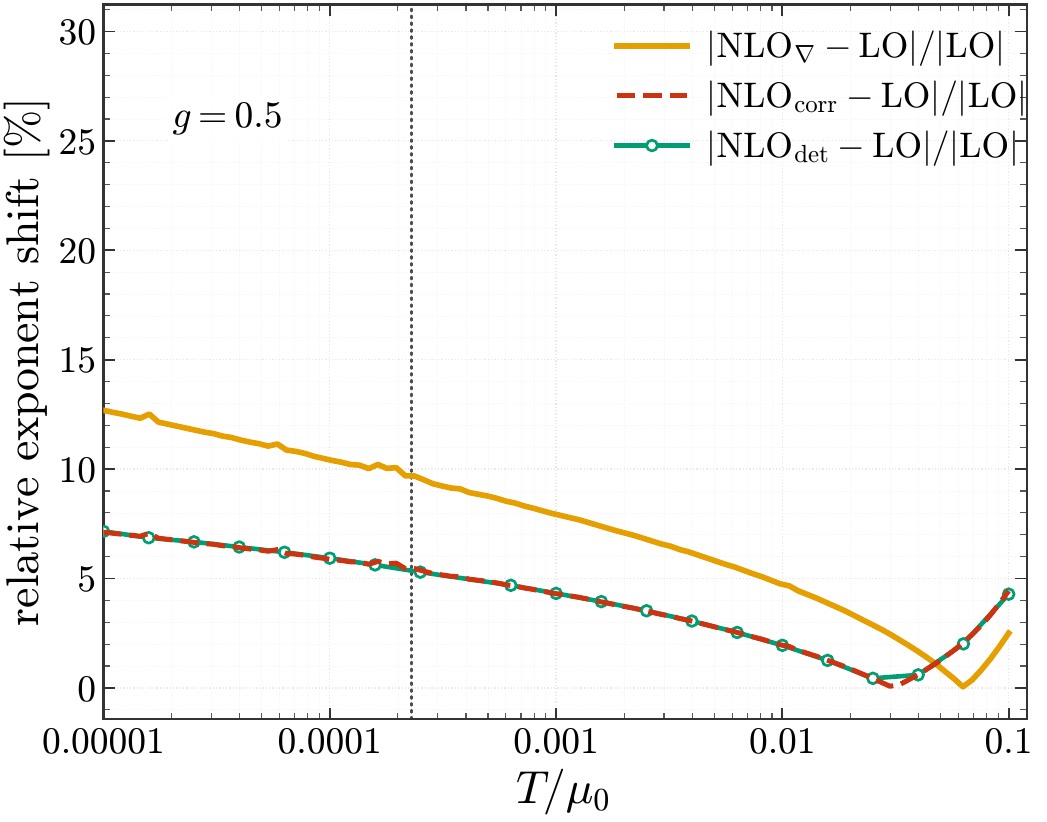}
        \caption{$g=0.5$: relative corrections.}
        \label{fig:actioncorr-relative-g05}
    \end{subfigure}
    \hfill
    \begin{subfigure}[t]{0.48\textwidth}
        \centering
        \includegraphics[width=\linewidth]{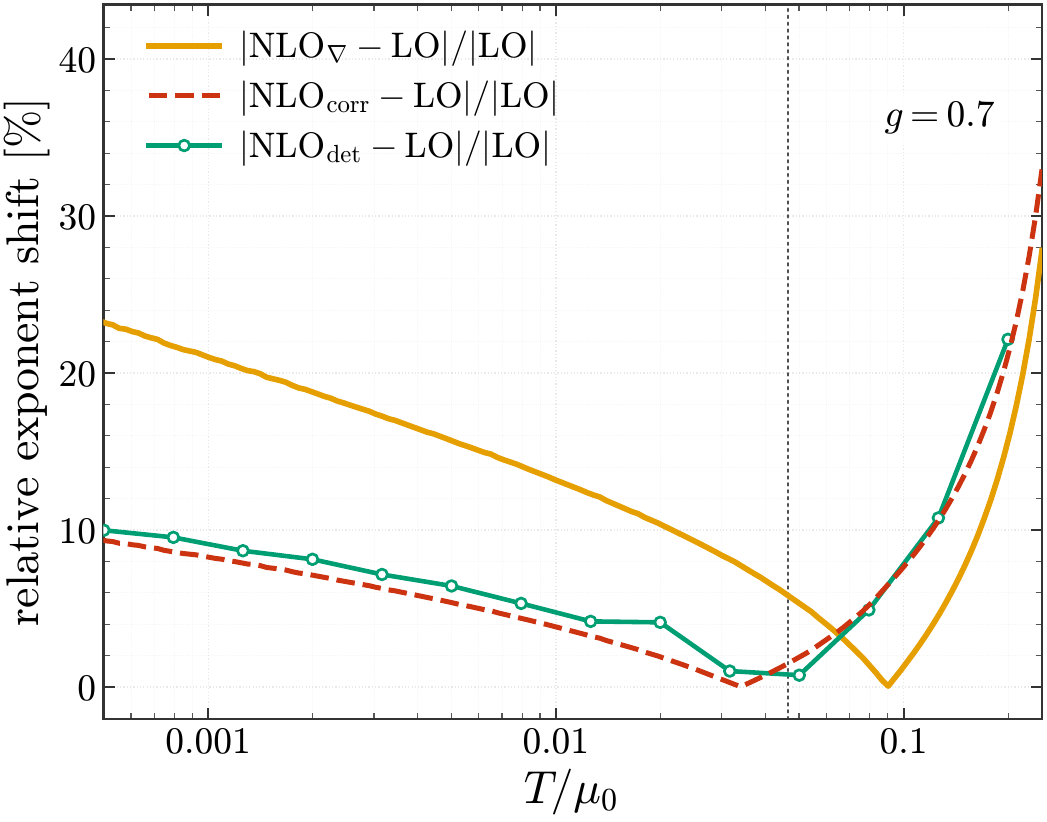}
        \caption{$g=0.7$: relative corrections.}
        \label{fig:actioncorr-relative-g07}
    \end{subfigure}

    \caption{
    Comparison of the LO, NLO gradient-expanded, corrected NLO, and
    functional-determinant results for $g=0.5$ and $g=0.7$.
    The upper panels show the nucleation-rate exponent, while the lower
    panels show the corresponding relative shifts with respect to LO.
    The vertical dotted lines indicate the percolation temperature of the
    corrected prescription for $\mu_0=1~\mathrm{GeV}$.
    }
    \label{fig:action-correct-factor}
\end{figure}

\Cref{fig:action-correct-factor} shows the resulting corrected prescription
for the two benchmark couplings. Around the percolation temperature, the
correction substantially improves the agreement with the functional
determinant, while retaining the simple local form of the gradient-expanded
calculation.

\subsection{Percolation and transition parameters}
\label{subsec:transition-parameters}

The nucleation of the first bubbles does not guarantee that the transition
completes. Percolation takes place once the expanding bubbles occupy a
sufficiently large fraction of space \cite{Athron:2022mmm,Levi:2022bzt,Matuszak:2026xsz}. As we are interested in transitions near the QCD scale, we furthermore should account for changes in the relativistic degrees of freedom. 

The entropy density is characterised by $g_{*s}(T)$, the effective number
of relativistic degrees of freedom contributing to the entropy density.
When $g_{*s}$ varies with temperature, entropy conservation modifies the
relation between temperature and cosmic time. For adiabatic expansion,
the comoving entropy is conserved~\cite{Husdal:2016haj}, such that
\begin{equation}
    s(T)a^3(T)=\mathrm{const.},
    \qquad
    a(T)\,T\,g_{*s}^{1/3}(T)=\mathrm{const.}
\end{equation}
Differentiating this relation gives the temperature--time evolution
\cite{Becker:2018rve},
\begin{equation}
    \frac{\mathrm dT}{\mathrm dt}
    =
    -H(T)T
    \left[
        1+
        \frac{T}{3g_{*s}(T)}
        \frac{\mathrm d g_{*s}(T)}{\mathrm dT}
    \right]^{-1}.
    \label{eq:temperature-time-relation}
\end{equation}
For compactness, we define
\begin{equation}
    \mathcal C_s(T)
    \equiv
    1+
    \frac{T}{3g_{*s}(T)}
    \frac{\mathrm d g_{*s}(T)}{\mathrm dT}
    =
    1+
    \frac{1}{3}
    \frac{\mathrm d\log g_{*s}(T)}
         {\mathrm d\log T}.
    \label{eq:Cs-definition}
\end{equation}
The probability for a point to remain in the false vacuum can then be
written as
\begin{equation}
\begin{aligned}
P_{\mathrm f}(T)
&=
e^{-I(T)},
\\[2pt]
I(T)
&=
\frac{4\pi v_w^3}{3}
\int_T^{T_c}
\mathrm dT'\,
\frac{\Gamma(T')\,\mathcal C_s(T')}
     {H(T')T'}
\frac{a^3(T')}{a^3(T)}
\\
&\qquad\times
\left[
\int_T^{T'}
\mathrm d\widetilde T\,
\frac{\mathcal C_s(\widetilde T)}
     {H(\widetilde T)\widetilde T}
\frac{a(T)}{a(\widetilde T)}
\right]^3 ,
\end{aligned}
\label{eq:false-vacuum-probability}
\end{equation}
which is the temperature-space percolation integral in~\cite{Matuszak:2026xsz}, written here using the compact notation
$\mathcal C_s(T)$ introduced above.

We define the percolation temperature by $P_{\mathrm f}(T_p)=0.71$,
corresponding to a converted volume fraction
$f_{\mathrm{perc}}=1-P_{\mathrm f}(T_p)=0.29$ \cite{Athron:2023xlk}.
Here $v_w$ denotes the bubble-wall velocity. We set $v_w=1$, appropriate
for the ultra-relativistic bubble walls expected in the strongly
supercooled regime under consideration. The integral is evaluated directly
using the full Hubble rate defined in \cref{eq:full-Hubble}. To speed up the scan over the dark gauge coupling $g$ and the input scale $\mu_0$, we use the saddle-point approximation of \cite{Rosauro-Alcaraz:2026kyj}  to identify the temperature interval containing $T_p$. The final value is obtained from the direct percolation integral.

For very slow transitions, particularly during vacuum domination, the
usual percolation criterion does not by itself guarantee that the physical
false-vacuum volume is already decreasing
\cite{Ellis:2018mja,Athron:2022mmm,Levi:2022bzt}.
Although $P_{\mathrm f}$ decreases as bubbles nucleate and grow, the
expansion of the Universe can compensate for this decrease. We therefore
consider the physical false-vacuum volume,
\begin{equation}
    \mathcal V_{\mathrm f}(t)
    \propto
    a^3(t)P_{\mathrm f}(t)
    =
    a^3(t)e^{-I(t)},
    \label{eq:physical-false-vacuum-volume}
\end{equation}
and evaluate the completion diagnostic
\begin{equation}
    \mathcal C_p
    \equiv
    \left.
    \frac{1}{H}
    \frac{\mathrm d\log\mathcal V_{\mathrm f}}
         {\mathrm dt}
    \right|_{T_p}.
    \label{eq:completion-diagnostic}
\end{equation}
The condition $\mathcal C_p<0$ implies that the physical false-vacuum
volume is already decreasing when percolation is reached, while
$\mathcal C_p=0$ marks the boundary between the two behaviours. For
$\mathcal C_p>0$, the false-vacuum volume is still growing at $T_p$.
This does not by itself exclude completion at a later time. We therefore
retain these points and indicate them separately in the parameter-space
plots as a completion diagnostic.

Having characterised both percolation and this completion diagnostic, we
proceed to compute the phase-transition parameters entering the
gravitational-wave spectrum.

Once it is ensured that the phase transition completes, we can proceed to compute the phase transition parameters that determine the GW spectrum. 
The strength parameter $\alpha$ compares the energy released during the transition with the radiation energy density. In the strongly supercooled regime, the broken minimum lies far from the field region controlling the thermal barrier. We use the dimensionally reduced potential to describe the barrier and compute the nucleation rate, but not to determine the depth of this distant minimum, which lies outside the range of validity of the thermal EFT. Thermal corrections are subdominant there, so the available vacuum energy is instead fixed by the zero-temperature potential \cite{Kierkla:2023von, Christiansen:2025xhv}. We therefore use

\begin{equation}
\alpha(T_p)=
\frac{\Delta V_{\mathrm{CW}}}{\rho_{\mathrm{rad}}(T_p)}
=
\frac{30\,\Delta V_{\mathrm{CW}}}
{\pi^2 g_{*\rho}(T_p)T_p^4},
\label{eq:alpha-definition}
\end{equation}

with $\Delta V_{\mathrm{CW}}$ given in \cref{eq:CW-vev}.

The inverse duration $\beta/H$ characterises how rapidly the transition
proceeds compared with the expansion of the Universe. It is commonly
obtained from the slope of the nucleation rate \cite{Caprini:2015zlo}. Using the
temperature--time relation in \cref{eq:temperature-time-relation}, this
gives
\begin{equation}
\frac{\beta_\Gamma}{H}
=
-\left.
\frac{T}{\mathcal C_s(T)}
\frac{\mathrm d\log\Gamma}{\mathrm dT}
\right|_{T_p}.
\label{eq:beta-rate-definition}
\end{equation}
For temperature-independent $g_{*s}$, $\mathcal C_s=1$, recovering the
familiar expression
\begin{equation}
\frac{\beta_\Gamma}{H}
=
-\left.
T\frac{\mathrm d\log\Gamma}{\mathrm dT}
\right|_{T_p}.
\end{equation}

For strongly supercooled transitions, however, the nucleation rate can
depart significantly from a simple exponential form, and the numerical
derivative is sensitive to small variations in the action
\cite{Megevand:2016lpr,Matuszak:2026xsz}. We instead determine the mean
bubble separation directly from the integrated bubble number density
\cite{Turner:1992tz,Matuszak:2026xsz},

\begin{equation}
\begin{aligned}
n_B(T_p)
&=
\int_{T_p}^{T_c}
\mathrm dT'\,
\frac{\Gamma(T')P_{\mathrm f}(T')\mathcal C_s(T')}
     {T'H(T')}
\left[
\frac{a(T')}{a(T_p)}
\right]^3 ,
\\[2pt]
R_\star
&=
n_B(T_p)^{-1/3}\,.
\end{aligned}
\label{eq:bubble-density-Rstar}
\end{equation}

For temperature-independent $g_{*s}$, one has
$\mathcal C_s=1$ and $aT=\mathrm{const.}$, and
\cref{eq:bubble-density-Rstar} reduces to the commonly used expression
\cite{Kierkla:2022odc}.

To provide the $\beta/H$ required by the gravitational-wave prescription, we match this numerical value of $R_\star$ to the exponential-nucleation relation at a finite converted fraction \cite{Megevand:2016lpr,Matuszak:2026xsz},

\begin{equation}
\left(\frac{\beta}{H}\right)_{R_\star,f_{\mathrm{perc}}}
=
\frac{(8\pi)^{1/3}v_w}
{H(T_p)R_\star f_{\mathrm{perc}}^{1/3}}\,.
\label{eq:beta-Rstar-mapping}
\end{equation}

The factor $f_{\mathrm{perc}}^{-1/3}$ accounts for evaluating the mean
bubble separation at the finite converted fraction reached at percolation,
rather than in the $f_{\mathrm{perc}}\to1$ limit.

Propagating both NLO treatments to percolation, the determinant result gives a lower $T_p$ and a larger $R_\star$ than the gradient prescription. The corresponding values of $\beta/H$ are approximately $4$--$12\%$ smaller across the benchmark points.
The corrected prescription~\ref{eq:corrected-gradient-action} substantially reduces these differences: across the determinant benchmarks it reproduces $T_p$ within $2.5\%$, $\alpha$ about $11\%$, and $\beta/H$ within $5\%$. In the strongly supercooled region, $g\leq0.7$, the residual difference in $\beta/H$ is below $1.7\%$.
The complete comparison is given in \cref{app:comparison-plots}. 

A larger correction arises from the mapping between $R_\star$ and
$\beta/H$. Since our percolation condition corresponds to
$f_{\mathrm{perc}}=1-P_{\mathrm f}(T_p)\simeq0.29$, omitting the factor
$f_{\mathrm{perc}}^{-1/3}$ lowers $\beta/H$ by approximately $34\%$
relative to the corrected value
\cite{Matuszak:2026xsz}. Note that the size of this discrepancy depends on the choice of the percolation criterion, and e.g. shrinks to 14\% for $P_f(T_p) = 1/e$ which is sometimes chosen in the literature.

Although the relative shifts in $T_p$ and $\alpha$ are larger, these quantities do not strongly affect the gravitational-wave spectrum in the regime under consideration, as we explain in the following. 

The vacuum energy released during the transition reheats the plasma.
Assuming that the released vacuum energy is promptly redistributed into the radiation bath of our benchmark cosmology, we define the reheating temperature $T_{\mathrm{rh}}$ as the temperature after the transition.
Energy conservation across the transition
gives~\cite{Bigazzi:2020avc}
\begin{equation}
\begin{aligned}
\rho_{\mathrm{rad}}^{\mathrm{br}}(T_{\mathrm{rh}})
&=
\rho_{\mathrm{rad}}^{\mathrm{sym}}(T_p)
+
\Delta V_{\mathrm{CW}},
\\[2pt]
g_{*\rho}^{\mathrm{br}}(T_{\mathrm{rh}})
T_{\mathrm{rh}}^4
&=
g_{*\rho}^{\mathrm{sym}}(T_p)T_p^4
+
\frac{30\,\Delta V_{\mathrm{CW}}}{\pi^2}.
\end{aligned}
\label{eq:reheating-temperature}
\end{equation}

Here $\Delta V_{\mathrm{CW}}$, defined in \cref{eq:CW-deltaV}, is the
vacuum-energy density released in the transition. The first term on the
right-hand side accounts for the radiation already present at
percolation, while the second accounts for the conversion of vacuum
energy into radiation during reheating. If $g_{*\rho}$ is taken to be constant and equal in the two phases,
\cref{eq:reheating-temperature} reduces to the familiar relation
$T_{\mathrm{rh}}^4
=
T_p^4
+
30\,\Delta V_{\mathrm{CW}}/(\pi^2g_*)
=
T_p^4(1+\alpha)$.
In our calculation, we instead retain the temperature and phase dependence of $g_{*\rho}$ and determine $T_{\mathrm{rh}}$ from \cref{eq:reheating-temperature}.
A more general treatment can follow the energy transfer and the thermal evolution of the two phases dynamically during the transition \cite{Matuszak:2026xsz}.

In the strongly supercooled regime,
$\Delta V_{\mathrm{CW}}\gg\rho_{\mathrm{rad}}^{\mathrm{sym}}(T_p)$.
The reheating temperature is therefore controlled mainly by the released
vacuum energy and by the broken-phase radiation content, rather than by
the precise value of $T_p$.
This explains why the comparatively large shifts in $T_p$ and $\alpha$
between the two nucleation prescriptions translate into a much smaller
change in the reheating temperature. Moreover,
$\alpha/(1+\alpha)\simeq1$ throughout the strongly supercooled region.
The dominant effect of higher order corrections on the gravitational-wave peak frequency and amplitude therefore comes
from the shift in $\beta/H$: 

\begin{equation}
f_{\mathrm{peak}}
\propto
T_{\mathrm{rh}}\frac{\beta}{H},
\qquad
\Omega_{\mathrm{GW,peak}}
\propto
\left(\frac{H}{\beta}\right)^2
\left(
\frac{\kappa\alpha}{1+\alpha}
\right)^2.
\label{eq:GW-parameter-scaling}
\end{equation}
For the more strongly supercooled benchmarks, $g\leq0.7$, the gradient
prescription predicts a peak frequency approximately $7$--$12\%$ higher
and a peak amplitude approximately $15$--$22\%$ lower than the
determinant calculation. 
With the corrected prescription, the remaining
differences in the transition parameters correspond, using
\cref{eq:GW-parameter-scaling}, to shifts below approximately $2\%$ in
the peak frequency and $4\%$ in the peak amplitude. At $g=0.8$, where
the transition is less strongly supercooled, these residual shifts increase
to approximately $5\%$ and $12\%$, respectively.
By comparison, omitting the $f_{\mathrm{perc}}$ correction would lower
$\beta/H$ by approximately $34\%$ throughout the parameter space.

Compared with our previous analysis~\cite{Christiansen:2025xhv}, the present treatment improves the
description of the phase transition at several stages. We compute the
nucleation rate consistently at NLO, rather than including higher-order
corrections only through the effective potential. We characterise the
transition at percolation rather than at nucleation, include the
temperature dependence of the radiation and entropy degrees of freedom
in the cosmological evolution, and determine $\beta/H$ from the
integrated bubble density and mean bubble separation. The comparison with the functional-determinant calculation further allows us to construct a corrected gradient prescription that closely reproduces the determinant benchmarks.
The residual difference is subleading compared with that associated with modeling the resulting gravitational-wave signal, discussed in the next section.

\section{Transition dynamics and gravitational-wave predictions}
\label{sec:GW-spectrum}

\subsection{Scale scan and transition parameters}
\label{subsec:scale-scan}

Having established the nucleation prescription, we now study the
transition across the two-dimensional parameter space of the model,
\begin{equation}
    0.4\leq g\leq0.95,
    \qquad
    10\,\MeV\leq\mu_0\leq50\,\GeV.
    \label{eq:scan-range}
\end{equation}
The scan uses the corrected NLO gradient prescription of \cref{eq:corrected-gradient-action}, without the dimension-six contribution.
The sampling is denser at low $g$, where the transition parameters vary most rapidly, while the range of scales is motivated by the frequency range of the GW signal observed by the PTA experiments. 

At each point in the scan, $\Delta V_{\rm CW}$ is fixed by the corresponding value of $\mu_0$ through \cref{eq:CW-deltaV}.
The classical scale invariance of the model leaves $\mu_0$ as the only mass scale in the tunnelling problem.
At fixed $g$, changing $\mu_0$ rescales all masses and temperatures by the same factor.
The dimensionless tunnelling action therefore remains unchanged when $x=T/\mu_0$ is held fixed.
The same action table can consequently be reused throughout the $\mu_0$ scan.
This scaling does not extend to the percolation condition, since the Hubble rate introduces the Planck scale and $\Gamma/H^4$ retains an explicit dependence on $\mu_0$.
Consequently, $T_p/\mu_0$ is not constant at fixed $g$. We therefore solve the percolation condition separately at every point and compute the corresponding $R_\star$, $\alpha$ and $\left(\beta/H_p\right)_{R_\star,f_{\rm perc}}$.
The rescaling relations and numerical implementation are described in \cref{app:CW-scaling,app:percolation-scan}.

\begin{figure}[t]
    \centering
    \includegraphics[width=0.85\textwidth]
    {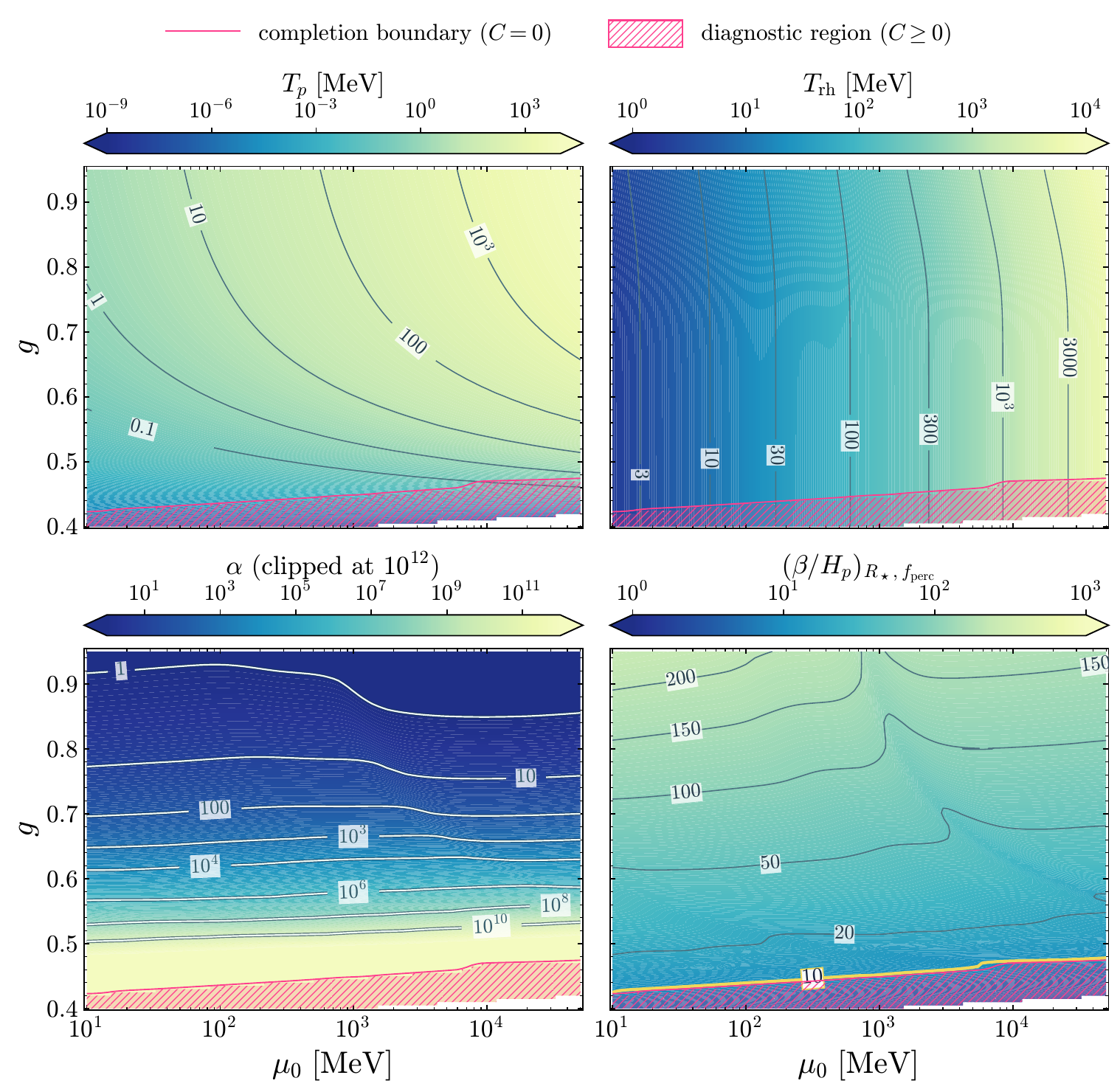}
    \caption{
        Phase-transition parameters across the $(g,\mu_0)$ plane:
        $T_p$, $T_{\rm rh}$, $\alpha$, and
        $(\beta/H)_{R_\star,f_{\rm perc}}$.
        The red contour marks $\mathcal C_p=0$, while the crossed region
        with $\mathcal C_p\geq0$ identifies points for which the physical
        false-vacuum volume is still growing at percolation.
    }
    \label{fig:transition-parameter-maps}
\end{figure}

Throughout the scan, we use the full expansion rate in
\cref{eq:full-Hubble} and set $v_w=1$ as above. The radiation and entropy
degrees of freedom are evaluated as temperature-dependent quantities.
For the Standard Model contribution we use the equation of state of
Ref.~\cite{Saikawa:2018rcs}, while the dark-sector contribution is
included explicitly in both phases. In the symmetric phase the complex
scalar and massless gauge field contribute four relativistic degrees of
freedom. In the broken phase their
contribution is instead evaluated with mass-dependent thermal weights
that account for the suppression of species as they become
non-relativistic~\cite{Matuszak:2026xsz}. We assume a common dark- and
visible-sector temperature throughout the
cosmological evolution for simplicity. The reheating temperature is
then determined from energy conservation using
\cref{eq:reheating-temperature}.

The qualitative behaviour in \cref{fig:transition-parameter-maps}
reflects the different roles of the two input parameters. At fixed
$\mu_0$, increasing $g$ raises $T_p/\mu_0$ and weakens the amount of
supercooling. Consequently, $\alpha$ decreases rapidly, whereas
$\left(\beta/H_p\right)_{R_\star,f_{\rm perc}}$ increases. At fixed
$g$, increasing $\mu_0$ generally raises both $T_p$ and $T_{\rm rh}$ in
physical units, but shifts the dimensionless percolation temperature
$x_p=T_p/\mu_0$ towards smaller values. This increases $\alpha$ and,
over most of the controlled region, decreases
$\left(\beta/H_p\right)_{R_\star,f_{\rm perc}}$.

For $g\gtrsim0.6$, the $\beta/H$ map also develops a visible valley as
$T_p$ passes through the QCD crossover region,
$T_p\sim100$--$200\,\mathrm{MeV}$. In this temperature range, the rapid
variation of $g_{*\rho}(T)$ and $g_{*s}(T)$ affects both the expansion
rate and the temperature--time relation, producing the structure seen
in \cref{fig:transition-parameter-maps}. In the vacuum-dominated limit,
$T_{\rm rh}$ is determined predominantly by
$\Delta V_{\rm CW}^{1/4}\propto\mu_0$ and therefore depends only weakly
on $g$.

Across the scan, we find percolation temperatures as low as $10^{-8}$~MeV, however the reheating temperatures remain mostly above 2~MeV, such that nucleosynthesis and neutrino decoupling are mostly unaffected.  The scan reaches values
$(\beta/H_p)_{R_\star,f_{\rm perc}}<10$ in the low-$g$ region, as
indicated explicitly by the corresponding contour in
\cref{fig:transition-parameter-maps}. Small values of $\beta/H$ are essential for a large GW signal in agreement with the PTA observations, however values below $10$ start to become problematic in itself. The smallest value of
$(\beta/H_p)_{R_\star,f_{\rm perc}}$ for which $\mathcal C_p<0$ is
approximately $9.3$. For slower transitions the
percolation criterion alone is not sufficient to establish that the
phase transition completes, and additional considerations are required, c.f.~\cref{subsec:transition-parameters}. These regions are hatched in red in the figures. Finally the tiny white region corresponds to points for which determining $T_p$ would require bounce solutions below the temperature range where our numerical calculation is under control, and therefore does not imply failure to percolate.

These transition parameters provide the input for the
gravitational-wave calculation discussed next.

\subsection{Gravitational-wave spectrum}
\label{subsec:GW-spectrum}

The mapping between the phase-transition parameters and the resulting
gravitational-wave spectrum introduces an additional theoretical
uncertainty, particularly for the strongly supercooled and
ultra-relativistic transitions considered here. In this regime, the
released vacuum energy can be distributed between the scalar-field
gradient and highly energetic plasma shells, and the subsequent
evolution of these sources after bubble collision is not yet fully
understood.

This issue was studied in detail in
\cite{Baldes:2024wuz}, where the conditions for
particle shells produced by relativistic bubble walls to free stream
until collision were derived. The parameter region relevant for our
model does not lie in the free-streaming regime identified there, since
interactions between the shell and the surrounding plasma become
important before collision. In this region, the subsequent evolution of
the relativistic shells, and consequently the resulting gravitational-wave
spectrum, is presently not known. In particular, the failure of the
free-streaming approximation does not by itself select a unique
hydrodynamic prescription.

For our central prediction we therefore adopt the prescription of
 \cite{Lewicki:2022pdb}.
This treatment follows separately the energy stored in the bubble wall and transferred to the plasma and provides numerical fits for the corresponding gravitational-wave contributions.
We use the dissipative behaviour $T_{rr}\propto R^{-3}$ for both the gauge-field bubble contribution and the relativistic fluid shell.
The same class of strong-transition gravitational-wave spectra is also adopted in \cite{Caprini:2024hue} for bubble collisions and highly relativistic fluid shells. 
This scaling is supported by the available numerical simulations for transitions breaking a gauge symmetry and for relativistic fluid shells \cite{Lewicki:2022pdb}.

The wall dynamics are determined from the pressure difference across the
bubble wall. We evaluate the leading and radiative friction pressures
following~\cite{Levi:2022bzt}, while the subsequent wall evolution
follows~\cite{Lewicki:2022pdb}. We use
\begin{align}
    \Delta P(\gamma)
    &=
    \Delta V
    -
    P_{\rm LO}
    -
    \widetilde P_{\rm NLO}\gamma\,,
    \nonumber\\
    \gamma_{\rm eq}
    &=
    \frac{\Delta V-P_{\rm LO}}
         {\widetilde P_{\rm NLO}}\,,
    \qquad
    R_{\rm eq}
    \simeq
    \frac{3}{2}R_0\gamma_{\rm eq}\,,
\label{eq:GW-wall-dynamics}
\end{align}
where $P_{\rm LO}$ and $\widetilde P_{\rm NLO}$ are evaluated using the
leading and radiative pressure contributions. The initial bubble radius \cite{Lewicki:2022pdb} is $R_0=2\sigma/\Delta P(\gamma=1)$, where $\sigma$ denotes the bubble-wall surface tension.
We evaluate $\sigma$ at $T_p$ from the LO
three-dimensional effective potential $\widehat V_3^{\mathrm{1L}}(\phi_3;T)$ used to construct the bubble,
with $\phi_4=\sqrt{T}\,\phi_3$ and $V_4=T V_3$. 
Following \cite{Lewicki:2022pdb}, we define the energy fraction
carried by relativistic bubble walls as
\begin{equation}
    \kappa_\phi(R)
    =
    \mathcal K
    \frac{R_{\rm eq}}{R}
    \frac{\gamma(R)}{\gamma_{\rm eq}},
    \qquad
    \mathcal K
    =
    \left(
        1-\frac{\alpha_\infty}{\alpha}
    \right)
    \left(
        1-\frac{1}{\gamma_{\rm eq}}
    \right),
\label{eq:GW-wall-efficiency}
\end{equation}
with $\alpha_\infty=P_{\rm LO}/\rho_{\rm rad}$. The factor
$\mathcal K$ accounts for the reduction of the energy available to
accelerate the wall once the leading friction contribution is included,
and we retain its full expression rather than approximating
$\mathcal K\simeq1$. The fluid efficiency is evaluated separately using the fluid-heating prescription of \cite{Lewicki:2022pdb}, rather than obtained it from the relation $\kappa_{\rm fl}=1-\kappa_\phi$.

An important consequence of this calculation is that $v_w=1$ should not be identified with a runaway wall.
Throughout the region of the two-dimensional scan $(g,\,\mu_0)$ where the wall-friction treatment is applicable, we find
\begin{equation}
    R_\star > R_{\rm eq},
\label{eq:wall-equilibrated}
\end{equation}
such that the wall reaches its equilibrium boost before the typical collision scale.
We find no valid scan point for which the wall is still accelerating at collision.

Towards larger gauge couplings, the wall-friction prescription becomes increasingly restricted, with the loss of valid solutions occurring first at small $\mu_0$. In the 2D scan, this starts around $g\simeq0.89$, while the affected region moves to larger $g$ as $\mu_0$ increases.
Since we are mainly interested in the strongly supercooled region, we do not investigate this further and remove the affected points from the gravitational-wave analysis. Close to this boundary, some of the remaining solutions also have $\gamma_{\rm eq}=\mathcal O(1)$, such that the large-$\gamma$ approximation used for the wall evolution becomes less reliable. 

Following \cite{Lewicki:2022pdb}, the present-day spectrum is
obtained by adding the bubble-wall and fluid contributions,
\begin{equation}
\begin{aligned}
h^2\Omega_{{\rm GW},0}(f)
={}&
1.67\times10^{-5}
\left(\frac{g_{*\rho}(T_{\rm rh})}{100}\right)
\left(\frac{100}{g_{*s}(T_{\rm rh})}\right)^{4/3}
\left(\frac{H_p}{\beta}\right)^2
\\
&\times
\sum_{i=\phi,\,{\rm fl}}
\left[
\frac{\kappa_i\alpha}{1+\alpha}
\right]^2
\mathcal S_i(f),
\end{aligned}
\label{eq:GW-spectrum-master}
\end{equation}
where the spectral shapes are parametrised as
\begin{equation}
\mathcal S_i(f)
=
A_i
\frac{(a_i+b_i)^{c_i}}
{
\left[
b_i
\left(\frac{f}{f_{p,i}}\right)^{-a_i/c_i}
+
a_i
\left(\frac{f}{f_{p,i}}\right)^{b_i/c_i}
\right]^{c_i}
}.
\label{eq:GW-spectral-shape}
\end{equation}
The corresponding present-day peak frequencies are obtained by
redshifting the simulation frequencies using the temperature-dependent
energy and entropy degrees of freedom~\cite{Saikawa:2018rcs},
\begin{equation}
\begin{aligned}
f_{p,i}
&=
\frac{h_*}{2\pi}
\left(\frac{\beta}{H_p}\right)
\left(\frac{2\pi f_p}{\beta}\right)_i,
\\
h_*
&=
1.65\times10^{-5}\,{\rm Hz}\,
\left(\frac{T_{\rm rh}}{100\,{\rm GeV}}\right)
\left(\frac{g_{*\rho}(T_{\rm rh})}{100}\right)^{1/2}
\left(\frac{100}{g_{*s}(T_{\rm rh})}\right)^{1/3}.
\end{aligned}
\label{eq:GW-peak-redshift}
\end{equation}
Here
$H_p/\beta\equiv
[(\beta/H_p)_{R_\star,f_{\rm perc}}]^{-1}$.
The parameters $A_i$, $a_i$, $b_i$, $c_i$, the peak locations
$(2\pi f_p/\beta)_i$, and the corresponding effective collision radii
are taken from Table~1 of Ref.~\cite{Lewicki:2022pdb}. The source
efficiencies are evaluated at these source-dependent effective collision
radii. Since the transition can be strongly supercooled, the frequency
redshift is evaluated using $T_{\rm rh}$ rather than $T_p$.
We also restore the causal infrared scaling
$\Omega_{\rm GW}\propto f^3$ below the redshifted Hubble frequency
$f_H=h_*/2\pi$ \cite{Caprini:2015zlo}.

The simulations underlying this prescription neglect the expansion of
the Universe during gravitational-wave production, an approximation
that requires $\beta/H\gg1$. More recently, \cite{Lewicki:2025hxg} studied the impact of cosmic expansion for slow transitions and found that it can significantly modify the
resulting spectrum. Since our scan contains points with small
$\beta/H$, we use this cosmic-expansion calculation as a comparison.

The cosmic-expansion calculation does not include the frictional wall dynamics relevant for our model and therefore does not impose the corresponding consistency conditions on the wall evolution. As a result, spectra can also be obtained at points where our central wall-friction prescription is no longer applicable. We do not interpret these additional points as an extension of the physical parameter region of the central prediction. A treatment consistently combining cosmic expansion, frictional energy transfer, and the subsequent evolution of the non-free-streaming plasma is currently unavailable.

\begin{figure}[t]
    \centering
    \includegraphics[width=0.88\textwidth]{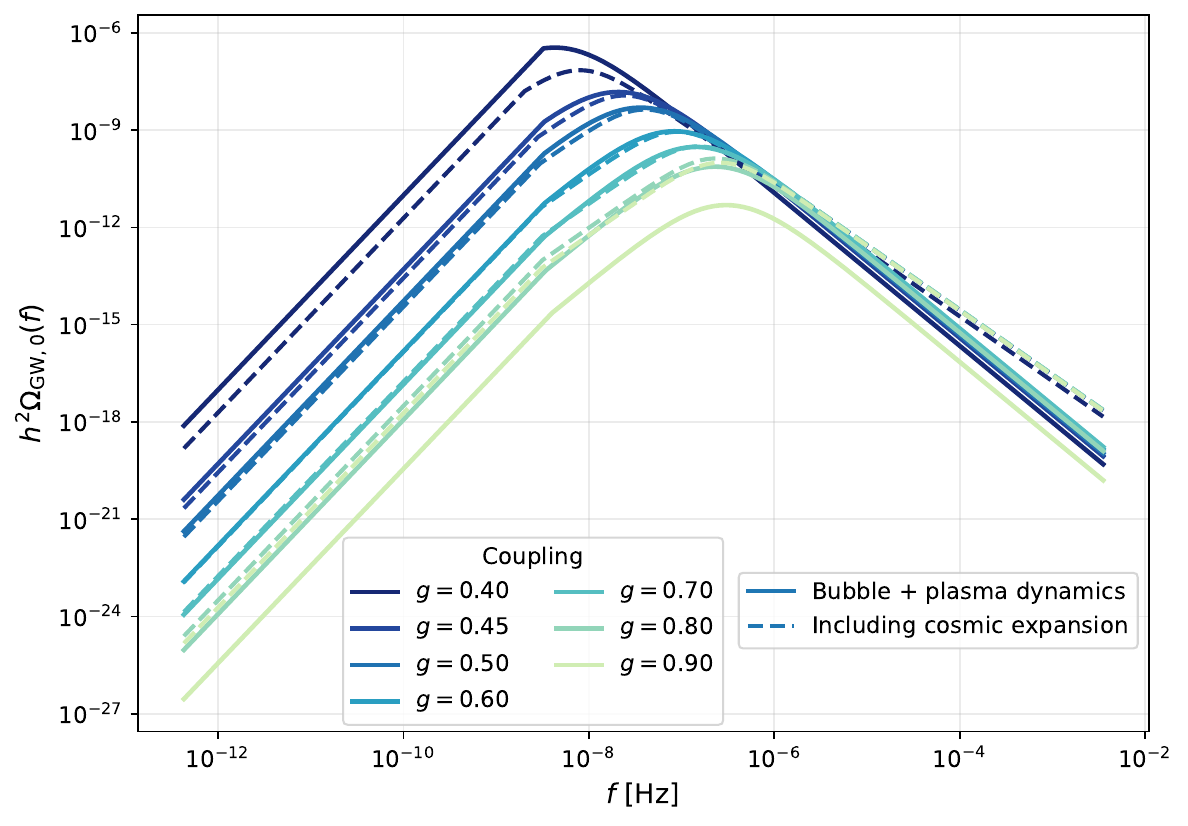}
    \caption{
    Present-day gravitational-wave spectra at $\mu_0=1\,{\rm GeV}$.
    Solid curves show our central bubble--fluid prediction based on
    \cite{Lewicki:2022pdb}, with the wall friction evaluated following
    \cite{Levi:2022bzt} and the dissipative
    $T_{rr}\propto R^{-3}$ source evolution. Dashed curves show the
    dissipative bulk-flow prescription including cosmic expansion
    from \cite{Lewicki:2025hxg}. 
    The transition parameters are obtained using the corrected NLO
    nucleation prescription.
}
    \label{fig:GW-cosmic-comparison}
\end{figure}

The direct comparison of the two prescriptions at
$\mu_0=1\,{\rm GeV}$ is shown in
\cref{fig:GW-cosmic-comparison}. For
$0.45\leq g\leq0.8$, the peak frequency is relatively stable, changing
by at most approximately $17\%$.
The largest shift occurs for the
slowest transition, $g=0.4$, where $\beta/H\simeq2$ and the peak
frequency increases by about $86\%$. The peak amplitude shows a somewhat
larger dependence on the prescription. For $0.45\leq g\leq0.7$, the two
predictions agree within roughly $20\%$, while larger differences appear
at the edges of the benchmark range. At $g=0.4$, the cosmic-expansion
result is approximately $80\%$ smaller, whereas at $g=0.9$ it exceeds the bubble--fluid prediction by more than an order of magnitude. For both prescriptions, the ultraviolet slope is independent of $g$ and is fixed by the corresponding spectral fit.
The cosmic-expansion prediction falls as $\Omega_{\rm GW}\propto f^{-2}$, while the central bubble--fluid prediction behaves approximately as $f^{-2.4}$.

The spectra shown here use the corrected gradient-expansion NLO nucleation prescription adopted for the main scan. As discussed in \cref{app:comparison-plots}, this substantially reduces the residual difference with the functional determinant benchmarks.
Using the scaling in \cref{eq:GW-parameter-scaling}, the remaining differences correspond to shifts below approximately $2\%$ in the peak frequency and $4\%$ in the peak amplitude for $g\leq0.7$, increasing to about $5\%$ and $12\%$, respectively, at $g=0.8$.
These residual differences are smaller than those associated with the gravitational-wave
prescription, illustrating that the modelling of the gravitational-wave source constitutes the larger theoretical uncertainty in this region.

\section{Phenomenology}
\label{sec:phenomenology}

We now confront the gravitational-wave spectra obtained above directly with PTA data.  Besides identifying the preferred region in $(g,\mu_0)$, this allows us to quantify how the theoretical treatment of the phase transition affects the inferred dark-sector parameters.

\paragraph{PTA likelihood and parameter inference.}
 We confront the predicted gravitational-wave spectra with the NANOGrav 15-year and IPTA DR2 free-spectrum results using \textsc{Ceffyl} \cite{Lamb:2023jls}.
For NANOGrav we use the first 14 Fourier modes, up to $f\simeq 27.7\,{\rm nHz}$, where most of the support for the common gravitational-wave signal is concentrated in the 15-year data \cite{NANOGrav:2023hvm}.
For IPTA DR2 we use the corresponding 25 modes up to $f\simeq 27.1\,{\rm nHz}$, so that both data sets are compared over approximately the same frequency interval.
For IPTA DR2, the free-spectrum reconstruction is provided at discrete Fourier frequencies \cite{Antoniadis:2022pcn,Lamb:2023jls}.
We therefore evaluate the predicted continuous gravitational-wave spectrum at these frequencies.
The apparent gaps between the IPTA intervals in the spectral figures reflect this discrete frequency sampling and not structure in the underlying gravitational-wave spectrum.
Technical details of the likelihood construction and the statistical treatment of the parameter grid are given in \cref{app:pta-likelihood} \cite{Antoniadis:2022pcn,Lamb:2023jls}.

\begin{figure}[t]
    \centering
    \includegraphics[
        width=0.78\textwidth
    ]{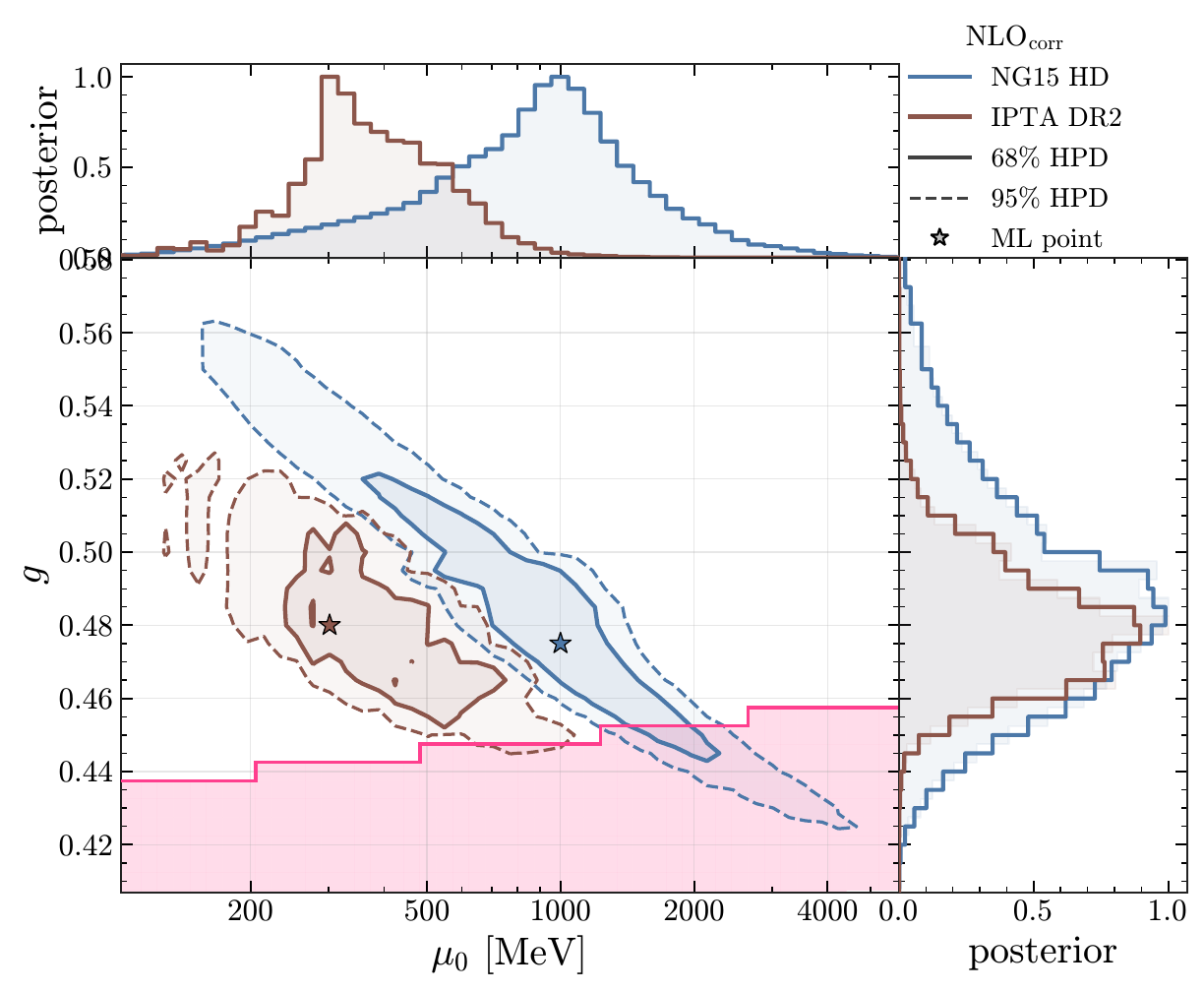}
    \caption{
        Posterior in the $(g,\mu_0)$ plane obtained using the corrected NLO
        gradient-expanded gravitational-wave templates.
        The blue and orange contours show the NANOGrav 15-year and
        IPTA DR2 results, respectively, with solid and dashed contours
        enclosing the 68\% and 95\% highest-posterior-density regions.
        Stars indicate the corresponding maximum-likelihood points.
        The $C_p=0$ contour separates the region in which the physical
        false-vacuum volume is already decreasing at percolation from
        the region with $C_p>0$.
        The one-dimensional marginalized posteriors are shown along the
        upper and right axes.
    }
    \label{fig:pta-ng15-ipta}
\end{figure}

IPTA DR2 combines the EPTA DR1, NANOGrav 9-year and PPTA DR1 data sets \cite{Antoniadis:2022pcn}.
It is therefore not statistically independent of the NANOGrav 15-year data.
We analyse the two likelihoods separately and do not construct a joint constraint.

The resulting NLO posteriors are shown in \cref{fig:pta-ng15-ipta}. 
The best-fit points are
\[
(g_{\rm ML},\mu_{0,\rm ML})=
\begin{cases}
(0.475,\,1.00~{\rm GeV}) & {\rm NG15},\\
(0.480,\,0.302~{\rm GeV}) & {\rm IPTA~DR2}.
\end{cases}
\]
Both best-fit points lie well inside the $C_p<0$ region, so the treatment of the $C_p>0$ region does not affect the preferred PTA solutions. 

The IPTA posterior is visibly more structured than the NG15 one, reflecting the different free-spectrum information entering the likelihood. 
The contours nevertheless overlap across the low-$g$ region, and the two best-fit points have very similar gauge couplings.
The main difference is instead the input scale $\mu_0$.
As discussed in \cref{sec:model}, $\mu_0$ fixes the overall scale of the radiatively generated potential, whereas $g$ controls its shape.
The two PTA data sets therefore select similar transition dynamics at different preferred physical scales.
This translates into reheating temperatures of approximately $150\,{\rm MeV}$ and $52.1\,{\rm MeV}$, and dark-photon masses of $1181\,{\rm MeV}$ and $356\,{\rm MeV}$, for NG15 and IPTA DR2, respectively. We note in passing that our best fit to IPTA has moved to smaller couplings, compared to~\cite{Madge:2023dxc} where the semi-analytic approximation of~\cite{Levi:2022bzt} was used to obtain the phase transition parameters. 

Both best-fit points nevertheless correspond to the same qualitative transition regime: strongly supercooled, relatively slow transitions, with $(\beta/H_p)_{R_\star,f_{\rm perc}}\simeq13.15$ for NG15 and $\simeq15.56$ for IPTA DR2.
Within the gravitational-wave scaling used here, such small inverse durations help place the signal in the PTA band by lowering the characteristic frequency and enhancing its amplitude.
The directly computed mean bubble separations are $H_pR_\star\simeq0.336$ and $0.284$, respectively.
For comparison, the NG15 best-fit value lies in the large $R_\star$ region preferred by the phenomenological NANOGrav bubble-collision analysis, $H_\star R_\star>0.28$ at 68\% credibility \cite{NANOGrav:2023hvm}.

At these values, cosmic expansion during gravitational-wave production is potentially relevant.
We quantified this effect in the previous section using the cosmic-expansion prescription of Ref.~\cite{Lewicki:2025hxg}.
As a further check, we repeat the PTA likelihood analysis on the subset of the scan for which spectra are available with both gravitational-wave prescriptions.
For NG15, the maximum-likelihood point shifts from $(g,\mu_0)=(0.475,1.00~{\rm GeV})$ to $(0.450,1.28~{\rm GeV})$, while for IPTA DR2 it shifts from $(0.480,0.302~{\rm GeV})$ to $(0.465,0.302~{\rm GeV})$.
Defining
$\Delta\ln\mathcal{L}_{\max}
=\ln\mathcal{L}_{\max}^{\rm cosmic}
-\ln\mathcal{L}_{\max}^{\rm pressure}$,
the corresponding changes in the maximum log-likelihood are
$+0.11$ and $-0.54$, respectively. Thus, the source prescription
shifts the preferred model parameters while giving a similar quality
of fit to the PTA data. Since the cosmic-expansion calculation does
not include the frictional wall dynamics relevant here, we treat the
difference between the two gravitational-wave prescriptions as a
source-model uncertainty.


\paragraph{Late reheating and nucleosynthesis.}
Both preferred PTA solutions correspond to very late percolation, $T_p\simeq39\,{\rm keV}$ for NG15 and $24.6\,{\rm keV}$ for IPTA DR2, followed by substantial reheating to $T_{\rm rh}\simeq150\,{\rm MeV}$ and $52\,{\rm MeV}$, respectively.
As in \cite{Madge:2023dxc}, we assume that the interaction transferring the released vacuum energy to the visible sector is sufficiently weak not to affect the phase-transition calculation.
If this transfer occurs promptly after the transition, both reheating temperatures are well above the few-MeV lower bound from BBN and neutrino thermalisation \cite{Bai:2021ibt,deSalas:2015glj}.

Since $T_p$ lies well below the MeV scale, the radiation bath crosses the usual nucleosynthesis temperature range before the transition and is reheated above it afterwards.
This is the same thermal history that motivated the two-stage nucleosynthesis discussion in \cite{Madge:2023dxc}.
The associated entropy production can also strongly dilute baryon or dark-matter abundances generated before the transition.

In the strictly secluded limit~\cite{Schwaller:2015tja,Breitbach:2018ddu,Fairbairn:2019xog}, the released energy would instead remain in the dark sector.
In that case, the single-temperature benchmark used here would be in conflict with constraints on $\Delta N_{\rm eff}$ in most of the parameter space. For a point to be viable, the dark sector energy density should always remain below that of the visible radiation sector. In the regime of large supercooling, this requires a tiny initial temperature ratio. A rough estimate is $T_D/T_{\rm SM} < T_p/\mu_0$. As discussed in~\cite{Bringmann:2023opz}, it is unlikely that such a scenario leads to a viable interpretation of the PTA signal. On the other hand a dark sector PT could still happen at different epochs, including times after nucleosynthesis, such as in models of early dark energy~\cite{Niedermann:2019olb}. The corresponding very low frequency GWs could e.g. lead to observable CMB spectral distortions~\cite{Kite:2020uix,Ramberg:2022irf}. 

\paragraph{Theory dependence of the inverse problem.}

A central question in current studies of cosmological phase transitions is how the theoretical treatment of the transition affects both the predicted gravitational-wave signal and its particle-physics interpretation.
We address this by comparing our corrected NLO result with a simpler one-parameter approximation (OPA) and repeating the PTA analysis using the corresponding gravitational-wave templates.
In the semi-analytic OPA treatment \cite{Levi:2022bzt}, the high-temperature potential can be rescaled such that the bounce action is determined by a universal function of a single dimensionless parameter.
This avoids solving the bounce numerically at every point in parameter space and makes the method particularly useful for large scans.
For the OPA grid used here, we include the running of the couplings and evaluate them at the renormalisation scale $\mu=\pi T$.
This choice is motivated by \cite{Christiansen:2025xhv}, where $\mu=\pi T$ was found to give good agreement between the four-dimensional high-temperature calculation and the dimensionally reduced result , and is different from our treatment in~\cite{Madge:2023dxc}, where the OPA was used but running was not included.
More recently, \cite{Rosauro-Alcaraz:2026kyj} performed a detailed comparison of semi-analytic high-temperature treatments with the full numerical 4D calculation.
At the same scale, they find that the running OPA reproduces the 4D result including Daisy resummation rather
well.
We therefore expect our comparison between OPA and the NLO treatment to closely approximate a comparison using the corresponding four-dimensional Daisy-resummed calculation  instead.

\begin{figure}[t]
    \centering
    \includegraphics[
        width=0.82\textwidth
    ]{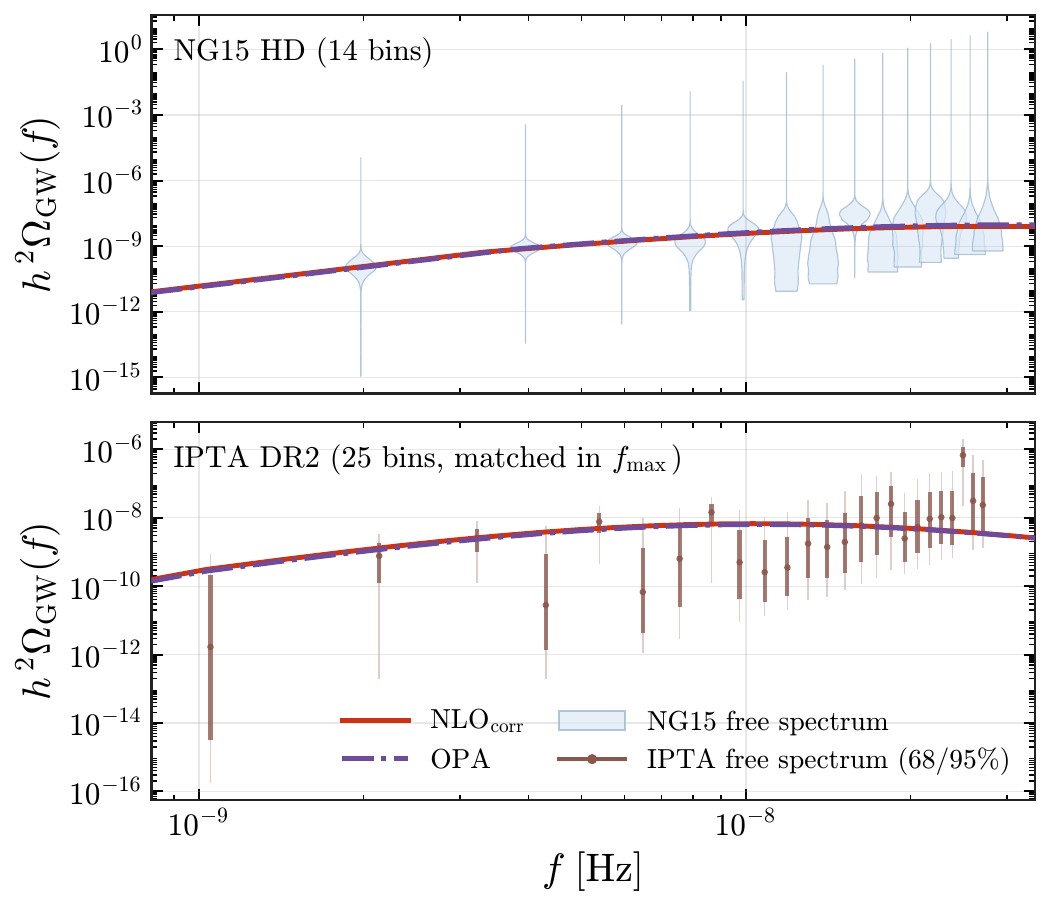}
    \caption{
        Best-fit gravitational-wave spectra for the corrected NLO gradient-expanded
        calculation and OPA, compared with the NANOGrav 15-year
        (upper panel) and IPTA DR2 (lower panel) free-spectrum results.
        The IPTA intervals correspond to the pointwise 68\% and 95\%
        ranges of the free-spectrum reconstruction.
    }
    \label{fig:pta-bestfit-spectra}
\end{figure}

The corresponding best-fit spectra are shown in
\cref{fig:pta-bestfit-spectra}. At fixed model parameters, the OPA and
NLO calculations can lead to sizeable differences in the phase-transition
dynamics and in the resulting gravitational-wave spectrum, particularly
towards small $g$, as illustrated in \cref{app:comparison-plots}.
A qualitative difference also appears in the strongly supercooled region.
For part of the low-$g$ parameter space, the percolation integral saturates
below the required threshold in the OPA calculation and no percolation
temperature can be assigned. We verified that this behaviour is unchanged
when extending the OPA evolution by many orders of magnitude towards lower
temperatures, and is therefore not set by the infrared boundary of the
numerical grid. The corresponding region is discussed in \cref{app:opa-no-percolation}.

Once the model parameters are fitted to the PTA data, however, the two
calculations give very similar spectra in the PTA band. The change in the
theoretical description is therefore compensated by a shift of the preferred
model parameters.

\begin{figure}[t]
    \centering
    \includegraphics[
        width=0.92\textwidth
    ]{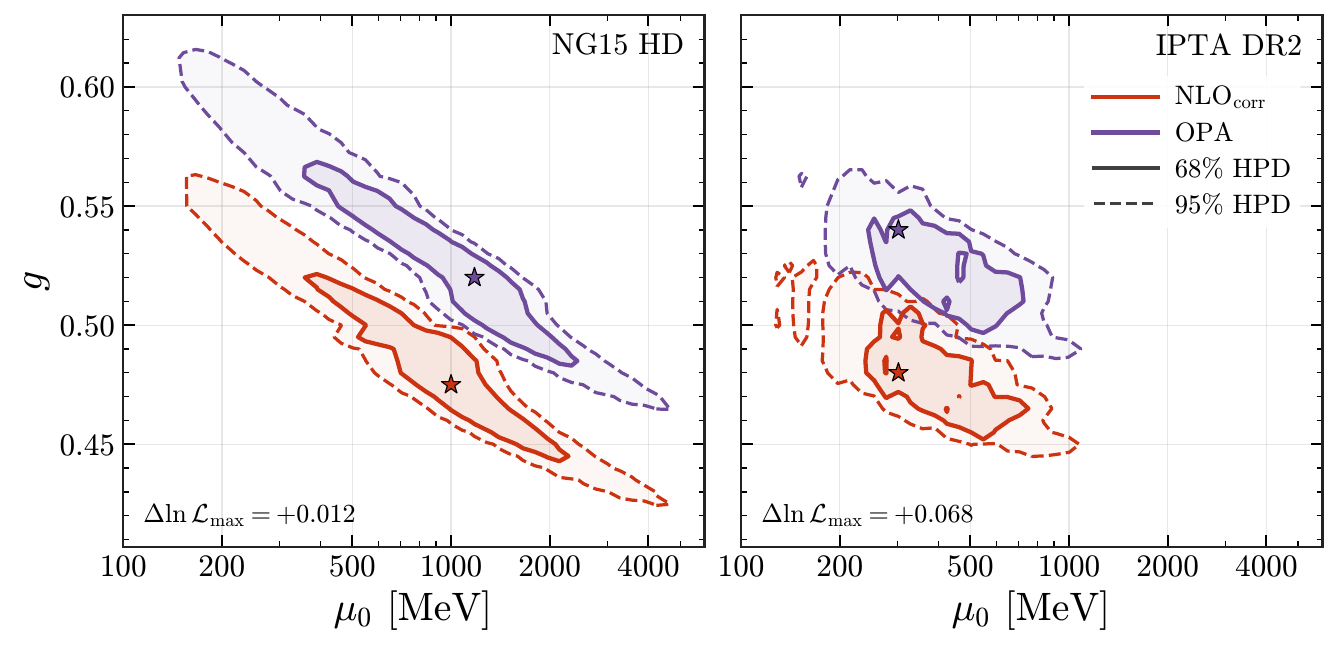}
    \caption{
        Dependence of the PTA parameter inference on the treatment of
        the phase transition.  The left and right panels show NG15 and
        IPTA DR2, respectively, comparing the corrected NLO gradient-expanded
        calculation with OPA.  Solid and dashed contours denote the
        68\% and 95\% HPD regions.
    }
    \label{fig:pta-theory-systematics}
\end{figure}

The close agreement of the best-fit spectra is reflected in the maximum
likelihood. Defining
\[
\Delta\ln\mathcal{L}_{\rm max}
\equiv
\ln\mathcal{L}_{\rm max}^{\rm OPA}
-
\ln\mathcal{L}_{\rm max}^{\rm NLO_{\rm corr}},
\]
we find
$\Delta\ln\mathcal{L}_{\rm max}=0.012$ for NG15 and $0.068$ for IPTA DR2.
This agreement, however, does not extend to the preferred particle-physics parameters.
As shown in
\cref{fig:pta-theory-systematics}, OPA shifts the preferred gauge
coupling from $g=0.475$ to $0.520$ for NG15 and from $g=0.480$ to
$0.540$ for IPTA DR2, corresponding to changes of approximately
$9\%$ and $13\%$, respectively.
For NG15, the preferred input scale also shifts from $\mu_0=1.00\,{\rm GeV}$ to $1.18\,{\rm GeV}$, while for IPTA DR2 it remains at $\mu_0\simeq0.302\,{\rm GeV}$.

\paragraph{Slow transitions and primordial black holes.}

The low-$g$ part of our NLO scan enters the regime of very slow
transitions that has recently been associated with primordial-black-hole
formation.  Statistical fluctuations in the nucleation history can leave
rare patches in the false vacuum for longer than the background, producing
large overdensities after the surrounding regions have transitioned
\cite{Lewicki:2023ioy,Gouttenoire:2023naa}.
In particular, \cite{Lewicki:2023ioy} finds the PBH abundance to be
controlled mainly by the mean bubble radius, with sizeable production for $H_p R_p\simeq0.5$; for an approximately exponential nucleation history this corresponds to $\beta/H\simeq3.8$.
Ref.~\cite{Gouttenoire:2023naa}
similarly finds observable PBH production for transitions with
$\beta/H\lesssim7$. 
A sizeable region of our scan reaches $H_pR_\star\gtrsim0.5$, while the slowest part of this region we find $(\beta/H_p)_{R_\star,f_{\rm perc}}\lesssim7$.
Since $R_\star$ is obtained directly from the integrated bubble number
density, it remains a well-defined measure of the transition length scale
also in the ultraslow regime, where the conventional action-derivative
definition of $\beta/H$ can become unreliable
\cite{Matuszak:2026xsz}.

Interestingly, this entire large-$R_\star$ (or equivalently small $\beta/H$) region lies on the $\mathcal C_p>0$ side of the completion boundary \cite{Athron:2022mmm} in \cref{fig:transition-parameter-maps}.
This means that, at our adopted percolation temperature $P_{\rm f}(T_p)=0.71$, the decrease of the false-vacuum fraction is not yet sufficient to overcome the expansion of the Universe, and the physical false-vacuum volume is still increasing.
It does not, however, imply that the transition cannot complete.
Following the evolution below $T_p$, the physical false-vacuum volume can start decreasing at a lower temperature $T_e$, defined by $\mathcal C(T_e)=0$, and subsequently reach a practical completion temperature $T_f$, which we define by $P_{\rm f}(T_f)=0.01$.
This delayed evolution occurs for the large majority of the points with $\mathcal C_p>0$ in our scan.
Most first reach $T_e<T_p$ and then continue to $T_f<T_e$. 
\footnote{Within the available transition history, 917 of the 950 points
with $\mathcal C_p>0$ reach a lower temperature $T_e<T_p$ at which
$\mathcal C(T_e)=0$, and subsequently reach
$P_{\rm f}(T_f)=0.01$. The remaining 33 points reach the lower limit of
our NLO action tables, $T/\mu_0\simeq5\times10^{-12}$, before the
corresponding criterion can be tested and are therefore left unresolved.} 
A similar issue was recently discussed in conformal gauge--Higgs models
in \cite{Kierkla:2025vwp}, where sufficiently slow transitions can
fail the shrinking-volume condition at $T_p$ while still completing at
later times.

The overlap between the PBH-relevant large-$R_\star$ regime and the
delayed evolution beyond $T_p$ makes this part of parameter space
particularly interesting.  A quantitative calculation would require
following the stochastic nucleation and reheating history of delayed
patches beyond $T_p$, together with a prescription for their subsequent
gravitational collapse.  Since the latter is itself under active
development, with recent covariant treatments substantially reducing
earlier estimates of the PBH abundance
\cite{Franciolini:2025ztf}, we do not impose a PBH constraint here.
A dedicated treatment of this region is left for future work.

\section{Conclusions}
\label{sec:conclusions}
In this work we have analysed the gravitational-wave phenomenology of the classically conformal Abelian Higgs model, with particular focus on the theoretical uncertainties relevant for strongly supercooled transitions.
We treat the nucleation rate consistently at NLO and use the functional-determinant calculation to test and improve the gradient expansion.
We then follow the transition through percolation and reheating, including the temperature dependence of the relativistic degrees of freedom, and determine the characteristic bubble scale from the nucleation history. Finally, we include the bubble-wall dynamics in the gravitational-wave prediction and study how the treatment of $\beta/H$ and of the gravitational-wave source affects the spectrum and its interpretation using PTA data.

Here, we demonstrate this effect directly in the inference from NG15 and IPTA DR2 and provide, to our knowledge, the first PTA analysis of a conformal dark-sector transition based on an NLO nucleation rate calibrated to the functional determinants. At the present PTA precision, higher-order corrections have been argued not to change the conclusion that the model can fit the signal~\cite{Bringmann:2026xcx}, while the determinant-level reconstruction study for LISA showed that they can affect its particle-physics interpretation~\cite{Kierkla:2026bnm}. Our results confirm both conclusions in the PTA context: replacing $\mathrm{NLO}_{\rm corr}$ with OPA shifts the preferred value of $g$ by approximately $9\%$ for NG15 and $13\%$ for IPTA DR2, and $\mu_0$ by $18\%$ for NG15, while their best-fitting spectra remain practically indistinguishable. The corresponding maximum log-likelihoods differ by only $0.012$ and $0.068$, respectively.

The functional determinants do not qualitatively change the predicted gravitational-wave signal, but their effect is not negligible. Across the benchmark range, $\mathrm{NLO}_{\nabla}$ predicts a peak frequency approximately $7$--$12\%$ higher and a peak amplitude $15$--$34\%$ lower than $\mathrm{NLO}_{\rm det}$. In the strongly supercooled regime, $\alpha/(1+\alpha)$ is already saturated and the reheating temperature is determined mainly by the vacuum energy. The larger differences in $T_p$ and $\alpha$ therefore translate into more moderate shifts in the spectrum, controlled primarily by the characteristic bubble scale and the inverse transition duration.

The corrected prescription reproduces the determinant nucleation exponent with a mean absolute relative difference of $0.8\%$ and reduces the maximum differences in $T_p$, $\alpha$, and $\beta/H$ to $2.5\%$, $11\%$, and $5\%$, respectively. For $g\leq0.7$, the remaining shifts in the gravitational-wave peak frequency and amplitude are below $2\%$ and $4\%$. The correction therefore allows the determinant result to be consistently propagated to the PTA analysis over the phenomenologically relevant region.

The limitations of OPA become qualitative at small coupling. For $g\lesssim0.41$, the percolation integral saturates before reaching the required value and no percolation temperature exists, even when the calculation is extended to $T/\mu_0=10^{-35}$. This is therefore a consequence of the transition dynamics predicted by OPA, rather than a numerical boundary effect.

Once the nucleation rate is calibrated to the determinants, its residual uncertainty is smaller than the changes produced by the gravitational-wave source prescription and by the mapping between $R_\star$ and $\beta/H$. For $g\leq0.7$, changing the source prescription shifts the peak frequency by up to $17\%$ and the amplitude by roughly $20\%$, comparable to the original gradient--determinant difference, as discussed in detail in Sec.~\ref{subsec:GW-spectrum}.

The condition $\mathcal C_p>0$ at percolation does not imply that the transition fails to complete. In most cases, $\mathcal C(T)$ becomes negative at a lower temperature $T_e<T_p$, after which the physical false-vacuum volume decreases and the transition continues to completion~\cite{Levi:2022bzt}. The PTA-preferred values, $\beta/H\simeq13$--$16$, overlap the range associated with primordial-black-hole production if delayed reheating generates an early matter-dominated era~\cite{Ai:2026zrs}, which may be not realized in realistic transitions however~\cite{Mansour:2026sdx}. Establishing whether the same region can generate an appreciable PBH abundance therefore requires a dedicated treatment of the delayed patches and the reheating history \cite{Flores:2024lng}.

If the PTA signal originates from a cosmological phase transition, connecting it to the symmetry-breaking scale and interactions of the dark sector requires control of both the nucleation rate and the gravitational-wave prediction.
Our analysis therefore provides not only an improved prediction for this model, but also a quantitative assessment of how reliably a nanohertz signal can constrain its particle-physics parameters.

\subsection*{Acknowledgements}
We would like to thank N.~Leister for carefully comparing fluctuation determinants with us and for numerous useful discussions and insights. 
Furthermore we thank
M.~Kierkla for sharing his fluctuation determinant code which we used for further comparisons and for discussion. We also thank B.~Swiezewska and P.~Schicho for useful discussions. CPI thanks Y.~Gouttenoire for useful insights in the gravitational spectrum treatment. The authors acknowledge support by the Cluster of Excellence ``PRISMA$^{++}$'' funded by the German Research Foundation (DFG) within the German Excellence Strategy (Project No. 390831469).
\clearpage
\appendix
\section{Model parameters and dimensionally reduced potential}
\label{app:model-equations}

We collect here the expressions required to reproduce the three-dimensional
potential. They were obtained with \texttt{DRalgo}~\cite{Ekstedt:2022bff} and
are reported in our conventions.
Writing $t=\log\mu$, the four-dimensional parameters run as
\begin{equation}
    \frac{\mathrm{d}g^2}{\mathrm{d}t}=\frac{g^4}{24\pi^2}\,,
    \qquad
    \frac{\mathrm{d}\lambda}{\mathrm{d}t}
    =\frac{3g^4-6g^2\lambda+10\lambda^2}{8\pi^2}\,,
    \qquad
    \frac{\mathrm{d}m^2}{\mathrm{d}t}
    =-\frac{m^2(3g^2-4\lambda)}{8\pi^2}\,.
    \label{eq:app-rges}
\end{equation}

\paragraph{Three-dimensional potential.} Quantities in the 3D EFT are indicated by a subscript, e.g. $g_3$. The matching relations to the 4D input parameters are given below. The potentials are related via
with 
\begin{equation}
    V_{3,\mathrm{tree}}(\phi_3)
    =-\frac{1}{2}m_3^2\phi_3^2+\frac{1}{4}\lambda_3\phi_3^4\,.
\end{equation}
The quantities $M_A$, $M_G$, $M_H$, and $M_0$ denote the field-dependent masses of the spatial gauge mode, the Goldstone mode, the radial scalar mode, and the temporal gauge mode $A_0$, respectively. In terms of the 3D parameters, they are given by
\begin{align}
    M_A^2&=g_3^2\phi_3^2\,,
    &M_G^2&=-m_3^2+\lambda_3\phi_3^2\,,
    &M_H^2&=-m_3^2+3\lambda_3\phi_3^2\,,
    &M_0^2&=m_D^2+\frac{1}{2}\lambda_{A\phi,3}\phi_3^2\,.
    \label{eq:app-tree-and-masses}
\end{align}
After subtracting the value at the origin, the one-loop potential is
\begin{align}
    \widehat V_3^{\mathrm{1L}}(\phi_3)
    &\equiv V_3^{\mathrm{1L}}(\phi_3)-V_3^{\mathrm{1L}}(0)\,,
    \nonumber\\[-2pt]
    V_3^{\mathrm{1L}}(\phi_3)
    &=V_{3,\mathrm{tree}}(\phi_3)
    -\frac{1}{12\pi}\operatorname{Re}
    \left(2M_A^3+M_G^3+M_H^3+M_0^3\right) .
    \label{eq:app-one-loop-potential}
\end{align}
At one loop, the roots are evaluated on their principal complex branches and only the real part of the resulting potential is retained. Consequently, a scalar mode with negative mass squared gives no real cubic contribution.\\

For the two-loop term, define 
\begin{align}
    x&\equiv\phi_3^2\,,
    \qquad\quad q\equiv g_3^2-3\lambda_3\,,\\
    \mathcal L(X)&\equiv\frac12+\log\!\frac{\mu_3}{X}\,,
    \nonumber\\[-2pt]
    K&\equiv(g_3^2)^2x+4\lambda_3^2x
    -2g_3^2(-2m_3^2+4\lambda_3x)\,,
    \nonumber\\[-2pt]
    C&\equiv\log(M_H+2M_A)-2\log(M_H+M_A)+\log M_H\,.
    \label{eq:app-two-loop-definitions}
\end{align}
The unnormalized result is
\begin{equation}
    \mathcal V_{3,\mathrm{2L}}
    =\mathcal V_0+\mathcal V_{AH}+\mathcal B_1
    +\mathcal B_{23}+\mathcal V_{\mathrm{rest}}\,,
    \label{eq:app-two-loop-decomposition}
\end{equation}
where
\begin{align}
    \mathcal V_0={}&
    \frac{g_3^2M_A(M_G+M_H)}{16\pi^2}
    +\frac{3\lambda_3(M_G^2+M_H^2)}{64\pi^2}
    +\frac{\lambda_3M_GM_H}{32\pi^2}
    -\frac{3x\lambda_3^2}{16\pi^2}\mathcal L(3M_H)\,,
    \nonumber\\[-2pt]
    \mathcal V_{AH}={}&\frac{(m_3^2)^2}{64\pi^2x}C\,,
    \qquad
    \lim_{\phi_3\to0}\frac{C}{4\phi_3^2}
    =-\frac{g_3^2}{4M_H^2(0)}\,,
    \label{eq:app-two-loop-first-blocks}
    \\
    \mathcal B_1={}&\frac14\Bigg\{
    \frac{(g_3^2)^2x}{8\pi^2}
    -\frac{g_3^2[m_3^2+(2g_3^2-3\lambda_3)x]}{16\pi^2}
    +\frac{g_3^2M_AM_H}{8\pi^2}
    \nonumber\\[-2pt]
    &-\frac{(-6m_3^2\lambda_3+9\lambda_3^2x)\mathcal L(M_H)}{16\pi^2}
    +\frac{(2m_3^2q+q^2x)\mathcal L(M_A+M_H)}{8\pi^2}
    \nonumber\\[-2pt]
    &-\frac{\{2m_3^2(2g_3^2-3\lambda_3)
    +x[q^2+7(g_3^2)^2-6g_3^2\lambda_3]\}
    \mathcal L(2M_A+M_H)}{16\pi^2}\Bigg\} ,
    \label{eq:app-two-loop-B1}
    \\
    \mathcal B_{23}={}&\frac14\Bigg\{
    \frac{-2M_A(g_3^2-2\lambda_3)M_G
    -2M_A(g_3^2+2\lambda_3)M_H+2g_3^2M_GM_H}{16\pi^2}
    \nonumber\\[-2pt]
    &+\frac{x\lambda_3^2\mathcal L(M_G+M_H)}{2\pi^2}
    -\frac{K\mathcal L(M_A+M_G+M_H)}{8\pi^2}\Bigg\} ,
    \label{eq:app-two-loop-B23}
    \\
    \mathcal V_{\mathrm{rest}}={}&
    -\frac{x\lambda_3^2\mathcal L(2M_G+M_H)}{16\pi^2}
    -\frac{x\lambda_{A\phi,3}^2\mathcal L(M_H+2M_0)}{64\pi^2}
    \nonumber\\[-2pt]
    &+\frac{\lambda_{A\phi,3}M_0(M_G+M_H)}{64\pi^2}
    +\frac{\lambda_{A,3}M_0^2}{128\pi^2}\,.
    \label{eq:app-two-loop-rest}
\end{align}
All roots and logarithms are evaluated on their principal complex branches;
negative mass squares are not replaced by absolute values inside logarithms of
sums. The contribution entering the action is
\begin{equation}
    \Delta V_{3,\mathrm{2L}}(\phi_3)
    =\operatorname{Re}\!\left[
      \mathcal V_{3,\mathrm{2L}}(\phi_3)-\mathcal V_{3,\mathrm{2L}}(0)\right] ,
    \qquad
    \widehat V_3^{\mathrm{2L}}
    =\widehat V_3^{\mathrm{1L}}+\Delta V_{3,\mathrm{2L}}\,.
    \label{eq:app-normalized-two-loop-potential}
\end{equation}

\paragraph{Matching relations.}
All four-dimensional parameters below are evaluated at $\mu_R$. We choose
$\mu_R=2\pi T$ and $\mu_3=g(\mu_R)T$, and define
$L_b=\log[\mu_R^2e^{2\gamma_E}/(16\pi^2T^2)]$ and
$c_+=(\gamma_E-L_b-12\log A_G)/2$, where
$A_G\simeq1.2824271291$. The matched couplings are
\begin{align}
    g_3^2={}&g^2T-\frac{g^4L_bT}{48\pi^2}\,,
    &\lambda_{A,3}={}&\frac{g^4T}{\pi^2}\,,
    \nonumber\\[-2pt]
    \lambda_3={}&\lambda T+\frac{T}{16\pi^2}
    [g^4(2-3L_b)+6g^2\lambda L_b-10\lambda^2L_b]\,,
    &\lambda_{A\phi,3}={}&2g^2T+\frac{g^2T}{24\pi^2}
    [24\lambda-g^2(L_b-4)]\,.
    \label{eq:app-matching-couplings}
\end{align}
The mass parameters are
\begin{align}
    m_D^2={}&\frac{g^2T^2}{3}
    +\frac{g^2}{144\pi^2}[12\lambda T^2+g^2(7-L_b)T^2-36m^2]\,,
    \nonumber\\[-2pt]
    m_3^2={}&m^2-\frac{T^2}{12}(3g^2+4\lambda)
    +\frac{L_b(3g^2-4\lambda)m^2}{16\pi^2}
    +\frac{g^4(8+216c_++39L_b)T^2}{576\pi^2}
    \nonumber\\[-2pt]
    &+\frac{\lambda^2(12c_++5L_b)T^2}{24\pi^2}
    -\frac{g^2\lambda(1+12c_++3L_b)T^2}{24\pi^2}
    \nonumber\\[-2pt]
    &-\frac{8(g_3^2)^2-16g_3^2\lambda_3+16\lambda_3^2
    +\lambda_{A\phi,3}^2}{32\pi^2}
    \log\!\frac{\mu_3}{\mu_R}\,.
    \label{eq:app-matching-masses}
\end{align}
The 3D parameters in the last line include their NLO matching corrections.

\paragraph{Gradient corrections.}
The field-dependent corrections to the scalar kinetic term are~\cite{Kierkla:2025vwp}
\begin{equation}
   Z_{\mathrm{sp}}(\phi_3)
   =
    -\frac{22g_3}{48\pi|\phi_3|},
    \qquad
    Z_0(\phi_3)
   =
    \frac{\lambda_{A\phi,3}^2\phi_3^2}
    {192\pi M_0^3(\phi_3)}.
    \label{eq:app-gradient-corrections}
\end{equation}

\paragraph{Determinant subtraction.}
To avoid counting twice the local gauge contribution already contained in
the leading bounce action, we include the subtraction factor
\begin{align}
\mathcal C_{\mathrm{sub}}
\bigl[\phi_b^{\mathrm{LO}}\bigr]
&=
-\frac{1}{12\pi}
\int \mathrm{d}^3x\,
\operatorname{Re}\Bigl[
2M_A^3\bigl(\phi_b^{\mathrm{LO}}\bigr)
+
M_0^3\bigl(\phi_b^{\mathrm{LO}}\bigr)
-
m_D^3
\nonumber\\
&\hspace{3.5cm}
+
\sum_{s=H,G}
\left(
M_s^3\bigl(\phi_b^{\mathrm{LO}}\bigr)
-
M_s^3(0)
\right)
\Bigr].
\label{eq:app-determinant-subtraction}
\end{align}
Compared with~\cite{Kierkla:2025qyz,Kierkla:2025vwp}, here in addition we have to include the scalar field contributions which are omitted from the LO action there. 

\paragraph{Dimension-six operator.}
After the \texttt{DRalgo} field redefinitions,
\begin{equation}
    \Delta V_{3,6}=c_6(\Phi_3^\dagger\Phi_3)^3
    =\frac{c_6}{8}\phi_3^6\,,
    \qquad
    c_6=\frac{\zeta(3)}{480\pi^4}
    \left(15g^6-18g^4\lambda+75g^2\lambda^2+100\lambda^3\right)_{\mu_R} .
    \label{eq:app-dimension-six}
\end{equation}
The one-loop potential determines the leading bounce; the two-loop and
dimension-six terms are evaluated perturbatively on that solution.

\section{Comparison of transition prescriptions}
\label{app:comparison-plots}
We compare four descriptions of the phase transition. Our central result, $\mathrm{NLO}_{\rm corr}$, is the determinant-calibrated correction to the NLO gradient expansion.
We also include the gradient result, $\mathrm{NLO}_{\nabla}$, discussed in \cref{subsec:nucleation-prescriptions}, to quantify the impact of the functional-determinant treatment on the transition parameters and the resulting gravitational-wave spectrum, and to verify the improvement obtained with the fitted correction.
The functional-determinant result, $\mathrm{NLO}_{\rm det}$, provides a benchmark in which the fluctuations are evaluated without a local derivative expansion. Finally, we consider the semi-analytic one-parameter approximation, OPA, in which the thermal potential is replaced by the high-temperature approximation used in earlier studies~\cite{Levi:2022bzt,Madge:2023dxc}.

Once the nucleation rate is specified, the subsequent cosmological evolution is treated consistently in all four calculations.
We use the same Hubble rate, temperature-dependent degrees of freedom, percolation condition, determination of $R_\star$, reheating prescription, and gravitational-wave model.
For $\mathrm{NLO}_{\nabla}$, $\mathrm{NLO}_{\rm corr}$, and OPA, we use the dimensional approximation $A(T)=T^4$.
In the determinant calculation, the dynamical prefactor and fluctuation determinants are evaluated explicitly around the LO bounce, as given in \cref{eq:determinant-rate}.
The gradient, corrected-gradient, and determinant calculations use the same wall surface tension, while for OPA it is evaluated from the OPA potential, as described in \cref{app:numerical-rescaling}.

The benchmark comparison is discussed in the main text.
Here we detail the construction of $\mathrm{NLO}_{\rm corr}$ and compare the four prescriptions across the coupling range, including their effect on the gravitational-wave spectrum.
We also discuss the loss of a percolation solution in OPA at small coupling and compare the different definitions of the inverse transition duration.

\subsection{Determinant-calibrated correction to the gradient expansion}
\label{app:gradient-correction}

We construct the corrected prescription using functional-determinant
calculations at five benchmark couplings, $g=\{0.45,\,0.50,\,0.60,\,0.70,\,0.80\}$, at $\mu_0=1\,\GeV$.
At each coupling--temperature point $(g_i,T_i)$, we define the comparison
exponent
\begin{equation}
    \mathcal E_{a,i}
    \equiv
    -\log\!\left[
        \frac{\Gamma_a(T_i;g_i)}{T_i^4}
    \right],
    \qquad
    a=\nabla,\mathrm{det},
\end{equation}
where $a$ labels the gradient and functional-determinant prescriptions.
Since the gradient calculation uses $A(T)=T^4$, its exponent is simply
$\mathcal E_{\nabla,i}=S_{\rm grad,NLO}(T_i;g_i)$. By contrast,
$\mathcal E_{{\rm det},i}$ also contains the explicitly evaluated
dynamical prefactor and fluctuation determinants.

Both calculations contain the same LO action and zero-momentum NLO potential correction, defined in \cref{eq:SLO,eq:DeltaSVNLO}.
To determine which gradient contribution best describes the remaining difference, we compare $\Delta\mathcal E/|S_{Z_{\rm sp}}|$ and $\Delta\mathcal E/|S_{Z_0}|$.
The first ratio is more nearly independent of temperature, with a mean absolute logarithmic slope of $0.045$, compared with $0.113$ for the temporal contribution.
This motivates a correction to the spatial gradient term.
This choice is also physically motivated, since the spatial gauge mode becomes light towards the symmetric-phase tail of
the bounce, where the gradient expansion loses control \cite{Kierkla:2025qyz}.
We therefore use the ansatz
\begin{equation}
    \Delta\mathcal E_i
    \simeq
    K A_i(p,q),
    \qquad
    A_i(p,q)
    =
    g_i^p x_i^q
    \left|S_{Z_{\rm sp},i}\right|,
    \qquad
    x_i=\frac{T_i}{\mu_0}.
    \label{eq:gradient-correction-ansatz}
\end{equation}
We determine $K$, $p$, and $q$ by minimising the squared residual over the full set of determinant points,
\begin{equation}
    \mathcal R(K;p,q)
    =
    \sum_i
    \left[
        K A_i(p,q)-\Delta\mathcal E_i
    \right]^2.
    \label{eq:gradient-correction-loss}
\end{equation}
For fixed $(p,q)$, the minimum in $K$ is analytic,
\begin{equation}
    K_{\min}(p,q)
    =
    \frac{
        \sum_i A_i(p,q)\Delta\mathcal E_i
    }{
        \sum_i A_i^2(p,q)
    }.
    \label{eq:gradient-correction-K}
\end{equation}
The global fit gives
\begin{equation}
    K=0.61695,
    \qquad
    p=1.36869,
    \qquad
    q=-0.03513.
    \label{eq:gradient-correction-fit}
\end{equation}
The small value of $q$ confirms that only a small additional temperature dependence is required.
Since $S_{Z_{\rm sp}}<0$ throughout the fitted range, the resulting correction can equivalently be written as
\begin{equation}
\begin{aligned}
    S_{\rm corr}
    &=
    S_{\rm grad,NLO}
    +
    K g^p x^q
    \left|S_{Z_{\rm sp}}\right|
    \\
    &=
    S_{\rm LO}
    +\Delta S_{V,\rm NLO}
    +S_{Z_0}
    +
    \left(1-Kg^p x^q\right)S_{Z_{\rm sp}},
\end{aligned}
\end{equation}
which is the form used in \cref{eq:corrected-gradient-action}.

Only the nucleation-rate exponent enters the fit.
The percolation observables therefore provide an independent test of the corrected prescription.
Across the fitted points, the corrected exponent reproduces the determinant result with a mean relative deviation of $0.80\%$.

\begin{figure}[t]
    \centering

    \begin{subfigure}[t]{0.47\textwidth}
        \centering
        \includegraphics[width=\linewidth]{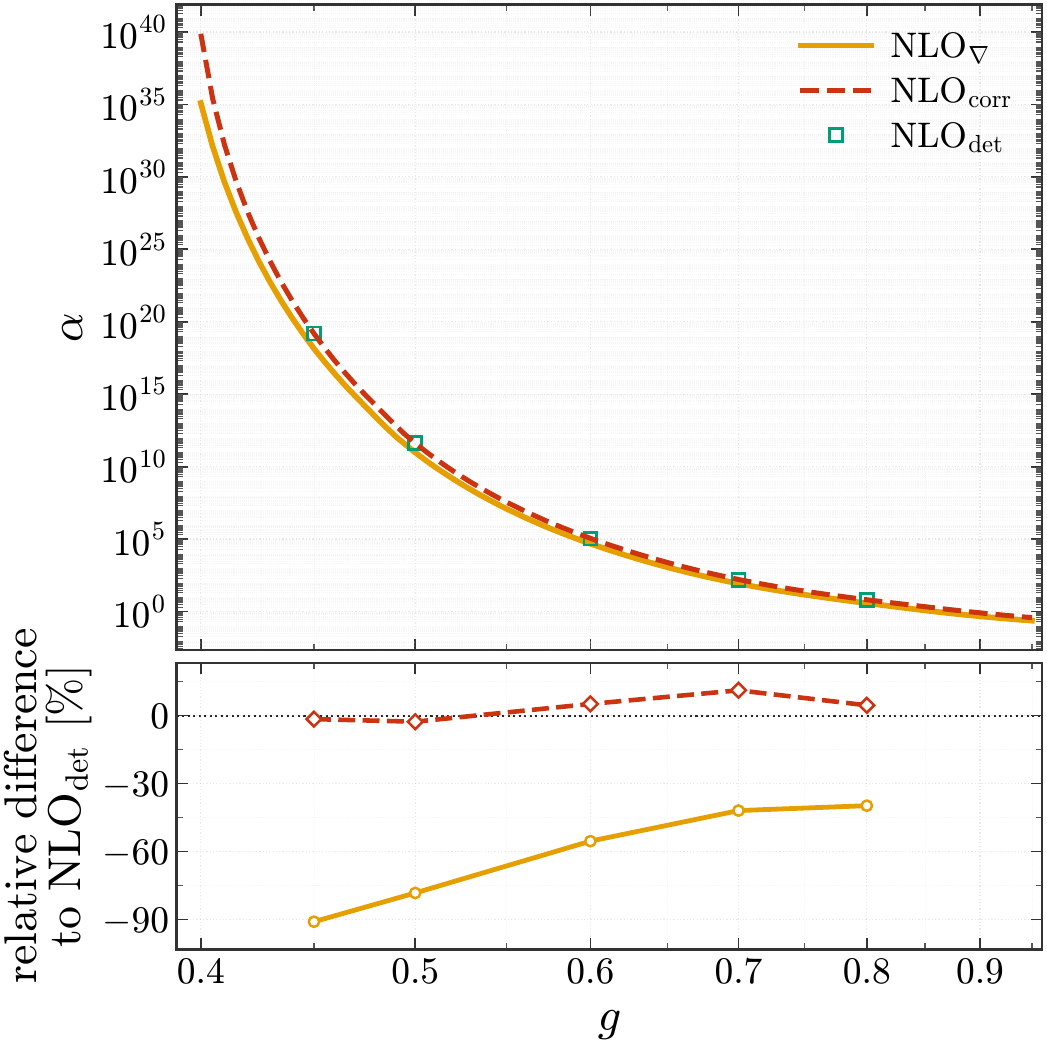}
        \caption{Transition strength.}
        \label{fig:corr-comparison-alpha}
    \end{subfigure}
    \hfill
    \begin{subfigure}[t]{0.47\textwidth}
        \centering
        \includegraphics[width=\linewidth]{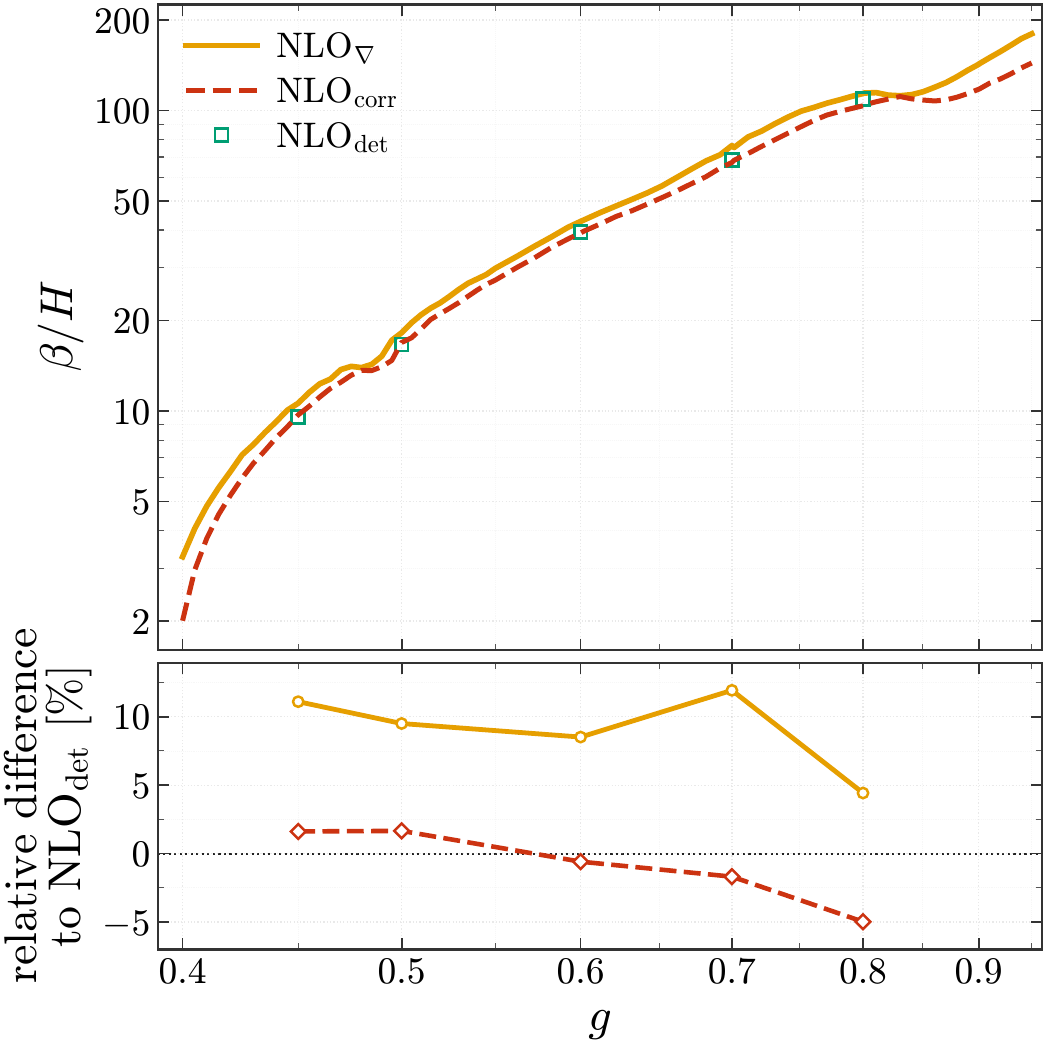}
        \caption{Inverse transition duration.}
        \label{fig:corr-comparison-betaH}
    \end{subfigure}

    \par\medskip

    \begin{subfigure}[t]{0.47\textwidth}
        \centering
        \includegraphics[width=\linewidth]{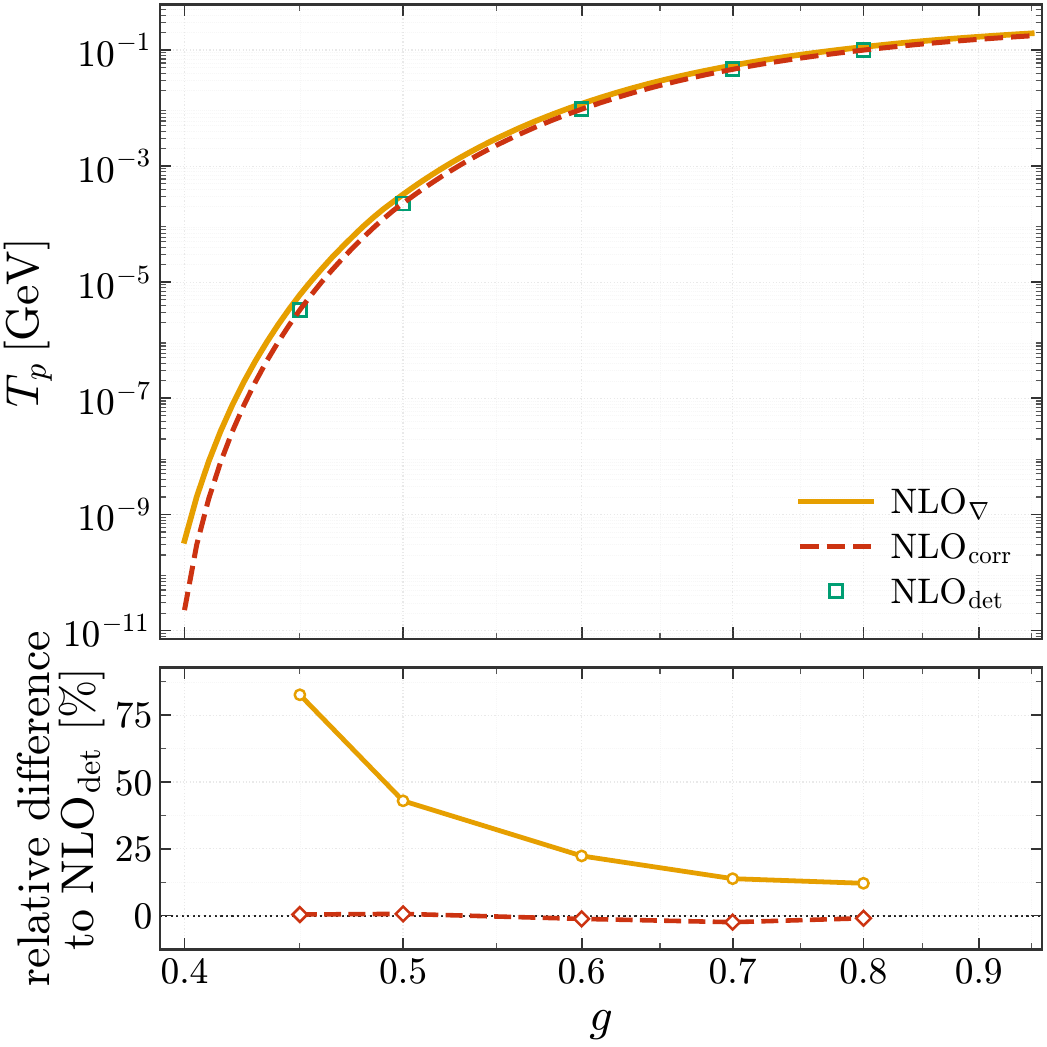}
        \caption{Percolation temperature.}
        \label{fig:corr-comparison-Tp}
    \end{subfigure}

    \caption{
        Comparison of the phase-transition parameters $T_p$,
        $\alpha$ and $\beta/H$ as functions of $g$ for
        $\mu_0=1\,\GeV$. The lines show the gradient
        ($\mathrm{NLO}_{\nabla}$) and corrected-gradient
        ($\mathrm{NLO}_{\rm corr}$) results, while the open squares denote the
        functional-determinant calculation ($\mathrm{NLO}_{\rm det}$). The lower
        panels show the signed relative differences of the two gradient
        prescriptions with respect to the determinant result,
        $100\,(X/X_{\rm det}-1)$.
}
    \label{fig:corr-comparison-observables}
\end{figure}

Propagating the corrected exponent through the percolation calculation provides a non-trivial test of the fit.
The transition parameters depend on different aspects of the full nucleation history, so improved agreement in the exponent does not automatically translate into a uniform improvement in $T_p$, $\alpha$, and $\beta/H$.
Relative to the determinant calculation,
the gradient prescription differs by $12.1$--$82.7\%$ in $T_p$,
$39.8$--$91.0\%$ in $\alpha$, and $4.4$--$11.9\%$ in $\beta/H$.
With the correction, the maximum differences are reduced to $2.5\%$, $11.2\%$, and $5.0\%$, respectively.
In the strongly supercooled region, $g\leq0.7$, the difference in $\beta/H$ remains below $1.7\%$.

We also tested several choices of $p$ and $q$ and propagated the resulting corrections through the percolation calculation. Among the forms considered, the prescription in \cref{eq:gradient-correction-fit} gives the best overall agreement with the determinant values of $T_p$, $\alpha$, and $\beta/H$ in the strongly supercooled region, $g\leq0.7$, which is the focus of this work.
Some alternatives give closer agreement at some of the larger-coupling benchmarks, but lead to larger deviations in at least one of these three quantities in the supercooled region. 

\subsection{Percolation observables across the coupling range}
\label{app:percobvs}

\begin{figure}[H]
    \centering

    \begin{subfigure}[t]{0.48\textwidth}
        \centering
        \includegraphics[width=\linewidth]{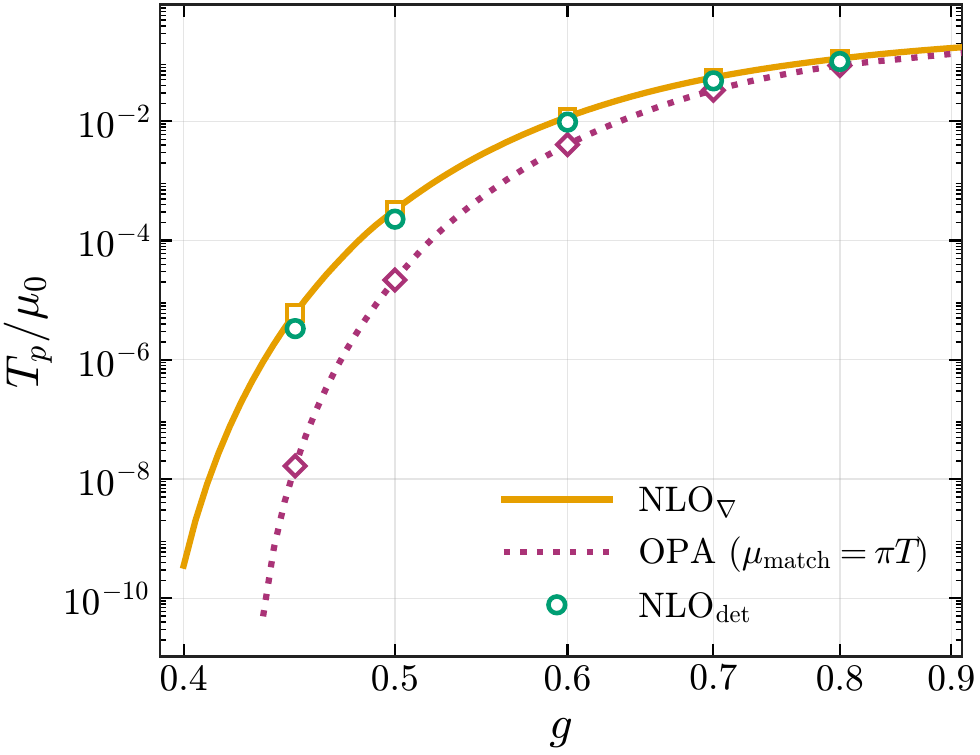}
        \caption{Percolation temperature.}
        \label{fig:Tp-det-grad}
    \end{subfigure}
    \hfill
    \begin{subfigure}[t]{0.48\textwidth}
        \centering
        \includegraphics[width=\linewidth]{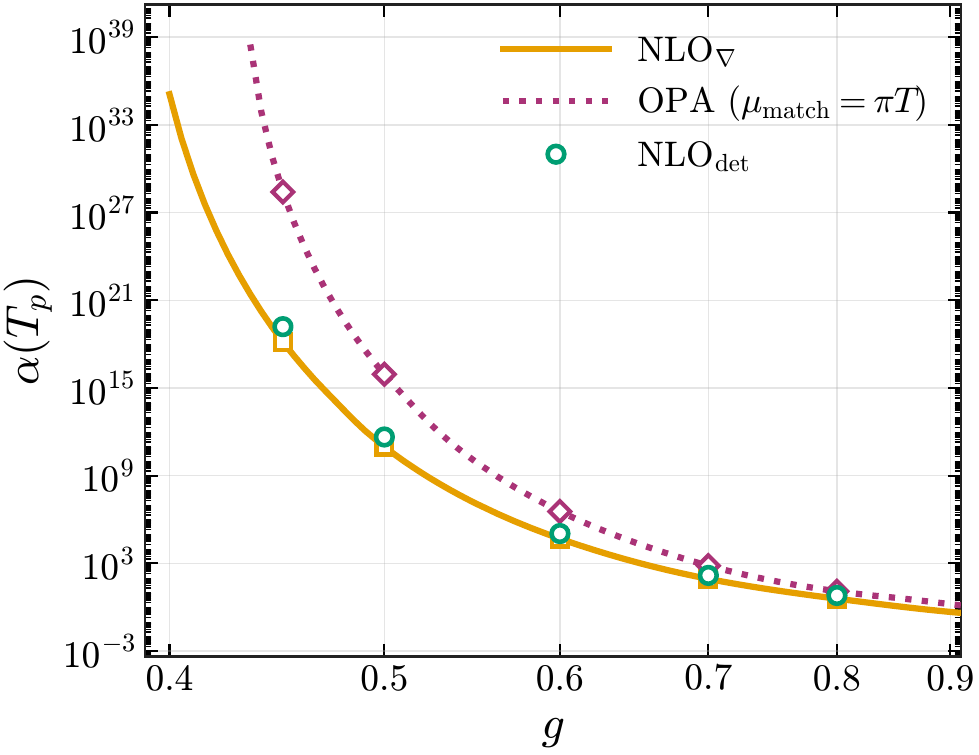}
        \caption{Transition strength.}
        \label{fig:alpha-det-grad}
    \end{subfigure}

    \par\medskip

    \begin{subfigure}[t]{0.48\textwidth}
        \centering
        \includegraphics[width=\linewidth]{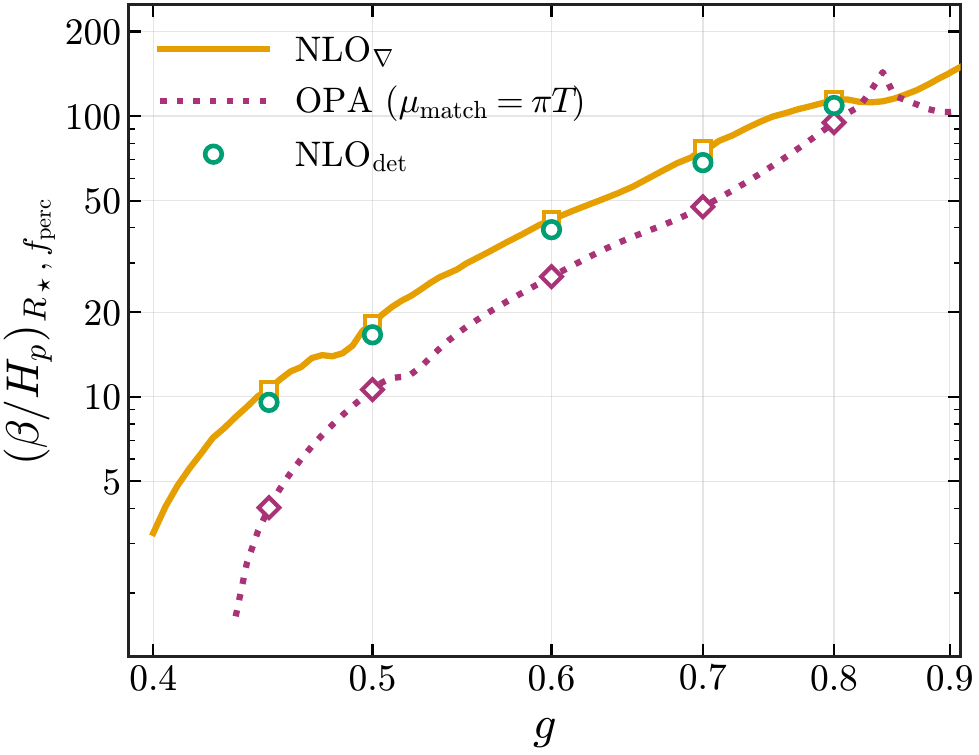}
        \caption{Inverse transition duration.}
        \label{fig:beta-det-grad}
    \end{subfigure}

    \caption{
        Percolation observables as functions of the gauge coupling $g$.
        The solid orange curves show the $\mathrm{NLO}_{\nabla}$ results, while the green
        circles denote the corresponding results obtained using the full
        fluctuation determinant: (a) $T_p/\mu_0$, (b) $\alpha(T_p)$, and
        (c) $(\beta/H)_{R_*,f_{\rm perc}}$.
    }
    \label{fig:percolation-det-grad}
\end{figure}

We furthermore compare our NLO determinant and NLO gradient approximated rates with the OPA approximation. The latter arguably offers the fastest numerical evaluation. 
\Cref{fig:percolation-det-grad} compares the three calculations at
$\mu_0=1\,\GeV$.
All three show the same qualitative dependence on the coupling, with stronger supercooling towards smaller $g$.
The differences between the prescriptions, however, become increasingly important in this region.

The difference between $\mathrm{NLO}_{\nabla}$ and OPA is the largest.
At $g=0.8$, the OPA percolation temperature is about $24\%$ below the
$\mathrm{NLO}_{\nabla}$ result. This difference increases to about $66\%$
at $g=0.6$, while at $g=0.45$ the OPA transition takes place more than two
orders of magnitude lower in temperature. The corresponding increase in
$\alpha$ is particularly large in the strongly supercooled regime, where the
radiation energy density at percolation is strongly suppressed. OPA also
gives a smaller inverse duration, with
$(\beta/H_p)_{R_\star,f_{\rm perc}}$ reduced by about $17\%$ at $g=0.8$
and $62\%$ at $g=0.45$.

The determinant correction is considerably smaller at the level of the
transition length and duration scales. The determinant calculation gives a
lower $T_p$ and a larger $\alpha$ than $\mathrm{NLO}_{\nabla}$, with the
difference increasing towards smaller $g$. At $g=0.45$, the relative shifts
in $T_p$ and $\alpha$ reach about $83\%$ and $91\%$, respectively, while at
$g=0.8$ they are about $12\%$ and $40\%$. In contrast,
$(\beta/H_p)_{R_\star,f_{\rm perc}}$ changes by at most about $12\%$ over
the benchmark range.
The difference between NLO and OPA is therefore mainly driven by the different finite temperature treatments. It is considerably larger than the change obtained when replacing the gradient expansion by the functional determinant.
\begin{table}[t]
    \centering
    \caption{
        Percolation parameters obtained with the NLO gradient expansion,
        the full fluctuation determinant and OPA for $\mu_0=1\,\mathrm{GeV}$.
        The relative deviations are defined as
        $\delta_X^{\rm det}
        =100\,|X_{\nabla}-X_{\rm det}|/|X_{\rm det}|$
        and
        $\delta_X^{\rm OPA}
        =100\,|X_{\rm OPA}-X_{\nabla}|/|X_{\nabla}|$.
    }
    \label{tab:determinant-percolation-comparison}

    \small
    \setlength{\tabcolsep}{5pt}
    \renewcommand{\arraystretch}{1.12}

    \begin{tabular}{@{}cccccc@{}}
        \toprule
        $g$
        & Prescription
        & $T_p/\mu_0$
        & $\alpha(T_p)$
        & $R_\star\,[\mathrm{GeV}^{-1}]$
        & $(\beta/H)_{R_\star,f_{\rm perc}}$
        \\
        \midrule

        $0.45$
        & $\mathrm{NLO}_{\nabla}$
        & \num{6.08e-6}
        & \num{1.39e18}
        & \num{2.59e19}
        & \num{10.6}
        \\
        &
        $\mathrm{NLO}_{\rm det}$
        & \num{3.33e-6}
        & \num{1.55e19}
        & \num{2.87e19}
        & \num{9.55}
        \\
        \rowcolor{gray!12}
        &
        \textbf{$\delta_X^{\rm det}\,[\%]$}
        & \bfseries\num{82.7}
        & \bfseries\num{91.0}
        & \bfseries\num{9.99}
        & \bfseries\num{11.1}
        \\
        &
        OPA
        & \num{1.65e-8}
        & \num{2.54e28}
        & \num{6.82e19}
        & \num{4.03}
        \\
        \rowcolor{gray!12}
        &
        \textbf{$\delta_X^{\rm OPA}\,[\%]$}
        & \bfseries\num{99.7}
        & \bfseries\num{1.83e12}
        & \bfseries\num{164}
        & \bfseries\num{62.1}
        \\
        \addlinespace

        $0.50$
        & $\mathrm{NLO}_{\nabla}$
        & \num{3.26e-4}
        & \num{9.46e10}
        & \num{1.51e19}
        & \num{18.2}
        \\
        &
        $\mathrm{NLO}_{\rm det}$
        & \num{2.28e-4}
        & \num{4.36e11}
        & \num{1.65e19}
        & \num{16.6}
        \\
        \rowcolor{gray!12}
        &
        \textbf{$\delta_X^{\rm det}\,[\%]$}
        & \bfseries\num{43.0}
        & \bfseries\num{78.3}
        & \bfseries\num{8.68}
        & \bfseries\num{9.50}
        \\
        &
        OPA
        & \num{2.18e-5}
        & \num{8.49e15}
        & \num{2.59e19}
        & \num{10.6}
        \\
        \rowcolor{gray!12}
        &
        \textbf{$\delta_X^{\rm OPA}\,[\%]$}
        & \bfseries\num{93.3}
        & \bfseries\num{8.97e6}
        & \bfseries\num{71.7}
        & \bfseries\num{41.7}
        \\
        \addlinespace

        $0.60$
        & $\mathrm{NLO}_{\nabla}$
        & \num{1.19e-2}
        & \num{4.80e4}
        & \num{6.42e18}
        & \num{42.7}
        \\
        &
        $\mathrm{NLO}_{\rm det}$
        & \num{9.70e-3}
        & \num{1.08e5}
        & \num{6.97e18}
        & \num{39.4}
        \\
        \rowcolor{gray!12}
        &
        \textbf{$\delta_X^{\rm det}\,[\%]$}
        & \bfseries\num{22.3}
        & \bfseries\num{55.4}
        & \bfseries\num{7.84}
        & \bfseries\num{8.51}
        \\
        &
        OPA
        & \num{4.06e-3}
        & \num{3.52e6}
        & \num{1.03e19}
        & \num{26.8}
        \\
        \rowcolor{gray!12}
        &
        \textbf{$\delta_X^{\rm OPA}\,[\%]$}
        & \bfseries\num{65.8}
        & \bfseries\num{7.23e3}
        & \bfseries\num{59.6}
        & \bfseries\num{37.3}
        \\
        \addlinespace

        $0.70$
        & $\mathrm{NLO}_{\nabla}$
        & \num{5.43e-2}
        & \num{8.54e1}
        & \num{3.57e18}
        & \num{76.4}
        \\
        &
        $\mathrm{NLO}_{\rm det}$
        & \num{4.77e-2}
        & \num{1.47e2}
        & \num{4.01e18}
        & \num{68.3}
        \\
        \rowcolor{gray!12}
        &
        \textbf{$\delta_X^{\rm det}\,[\%]$}
        & \bfseries\num{13.8}
        & \bfseries\num{42.0}
        & \bfseries\num{10.9}
        & \bfseries\num{11.9}
        \\
        &
        OPA
        & \num{3.32e-2}
        & \num{6.86e2}
        & \num{5.77e18}
        & \num{47.5}
        \\
        \rowcolor{gray!12}
        &
        \textbf{$\delta_X^{\rm OPA}\,[\%]$}
        & \bfseries\num{39.0}
        & \bfseries\num{704}
        & \bfseries\num{61.7}
        & \bfseries\num{37.8}
        \\
        \addlinespace

        $0.80$
        & $\mathrm{NLO}_{\nabla}$
        & \num{1.13e-1}
        & \num{3.76}
        & \num{2.14e18}
        & \num{114}
        \\
        &
        $\mathrm{NLO}_{\rm det}$
        & \num{1.01e-1}
        & \num{6.24}
        & \num{2.34e18}
        & \num{109}
        \\
        \rowcolor{gray!12}
        &
        \textbf{$\delta_X^{\rm det}\,[\%]$}
        & \bfseries\num{12.1}
        & \bfseries\num{39.8}
        & \bfseries\num{8.33}
        & \bfseries\num{4.42}
        \\
        &
        OPA
        & \num{8.63e-2}
        & \num{1.22e1}
        & \num{2.78e18}
        & \num{94.8}
        \\
        \rowcolor{gray!12}
        &
        \textbf{$\delta_X^{\rm OPA}\,[\%]$}
        & \bfseries\num{23.6}
        & \bfseries\num{224}
        & \bfseries\num{29.9}
        & \bfseries\num{16.7}
        \\

        \bottomrule
    \end{tabular}
\end{table}

\Cref{tab:determinant-percolation-comparison} makes the different size of the
two comparisons explicit. For the determinant calculation, the relative
difference in $T_p$ and $\alpha$ increases towards small coupling. At
$g=0.45$ the shifts are large, even though the corresponding change in
$(\beta/H_p){R\star,f_{\rm perc}}$ is only about $11\%$. The same pattern
continues over the benchmark range, with the determinant result remaining
close to the gradient calculation at the level of the transition length and
duration scales.

\begin{figure}[t]
    \centering
    \includegraphics[width=\linewidth]{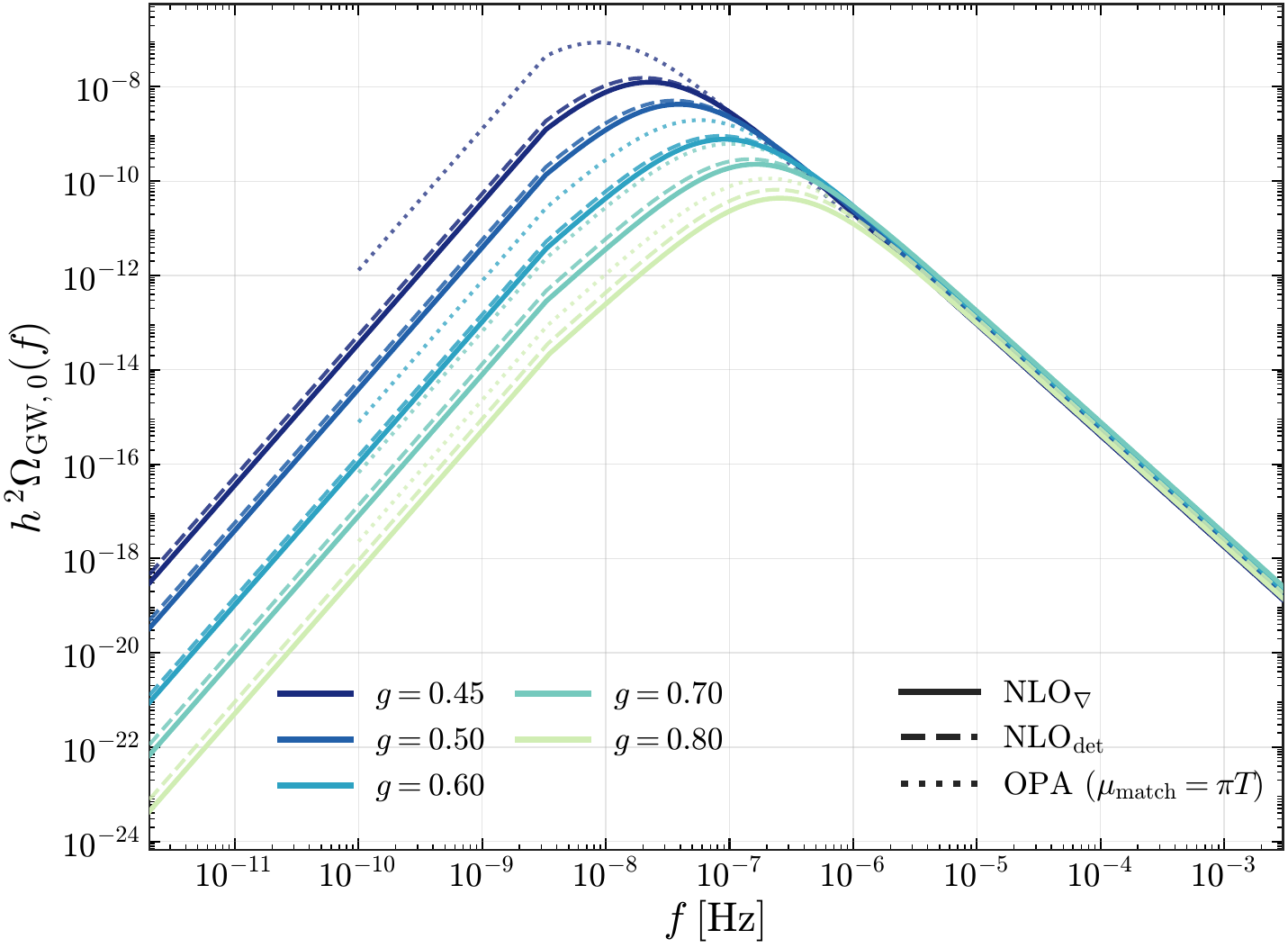}
    \caption{
        Gravitational-wave spectra for the benchmark couplings at
        $\mu_0=1\,\GeV$. Solid, dashed and dotted curves correspond to
        $\mathrm{NLO}_{\nabla}$, $\mathrm{NLO}_{\rm det}$ and OPA,
        respectively. The same bubble--fluid spectral prescription is used
        in all three calculations.
    }
    \label{fig:GW-grad-vs-det}
\end{figure}

\Cref{fig:GW-grad-vs-det} shows the corresponding gravitational-wave
spectra. The bubble and fluid spectral functions are the same in all three
calculations. The differences arise from the phase-transition parameters
and, in the OPA calculation, from the surface tension entering the wall
dynamics.

Across the benchmark range, $\mathrm{NLO}_{\nabla}$ gives a peak
frequency approximately $7$--$12\%$ higher and a peak amplitude
approximately $15$--$34\%$ lower than $\mathrm{NLO}_{\rm det}$.
The moderate shift in the peak frequency is consistent with the comparatively small differences in $R_\star$ and $(\beta/H_p)_{R_\star,f_{\rm perc}}$ shown in \cref{tab:determinant-percolation-comparison}.
In the strongly supercooled regime, where $\alpha$ is already large, these quantities control the remaining variation in the gravitational-wave spectrum.
Thus, using the gradient expansion instead of the functional determinants does not qualitatively change the predicted signal, but introduces a tens-of-percent uncertainty in its normalization.

The OPA calculation produces substantially larger shifts. Over the
benchmark range $0.45\leq g\leq0.8$, the peak frequency is approximately
$20$--$62\%$ lower than for $\mathrm{NLO}_{\nabla}$, while the peak
amplitude is larger by a factor of approximately $2.5$--$6.9$.
The largest difference occurs at $g=0.45$, the lowest coupling for which the OPA transition still percolates at $\mu_0=1\,\GeV$.

\subsection{Absence of percolation in OPA at small coupling}
\label{app:opa-no-percolation}

The OPA calculation develops a non-percolating region at small coupling.
On the $75\times102$ grid defined in \cref{app:numerical-rescaling},
$652$ OPA points do not satisfy the percolation condition in
\cref{eq:false-vacuum-probability}. For $g=0.400$ and $g=0.405$, none of
the values of $\mu_0$ in the scan percolate. At $g=0.41$, a solution is
found for $10$ of the $102$ values of $\mu_0$, increasing to $92$ at
$g=0.45$. The full $\mu_0$ range percolates from $g=0.46$ onwards. This
accounts for the termination of the OPA curves in
\cref{fig:percolation-det-grad}.

Extending the OPA calculation to much lower temperatures does not change
this result: the percolation integral saturates below $I_p=-\log(0.71)$. Within the OPA treatment, these points therefore do not percolate.
 The effect appears
in the same small-$g$ region where OPA predicts the strongest supercooling,
and represents a qualitative difference with respect to the NLO
calculation.

\subsection{Definitions of the inverse duration}
\label{app:beta-prescriptions}

\begin{figure}[t]
    \centering
    \includegraphics[width=0.8\textwidth]
    {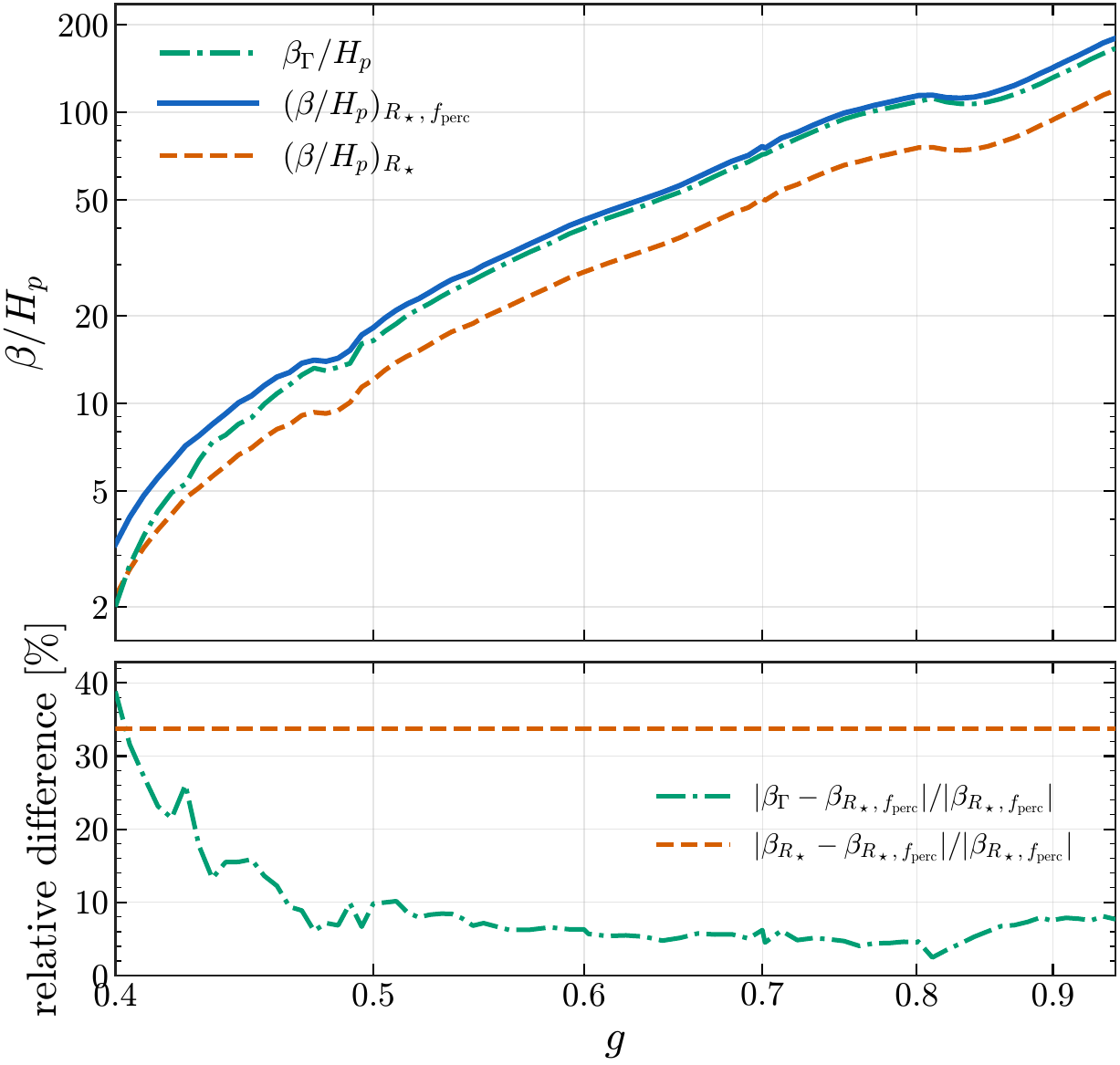}
    \caption{
        Inverse transition duration at $\mu_0=1\,\GeV$.
        The local rate derivative $\beta_\Gamma/H_p$ is shown in green.
        The blue curve shows our central result,
        $(\beta/H_p)_{R_\star,f_{\rm perc}}$, obtained from the mean bubble
        separation including the finite converted fraction at percolation.
        The orange curve shows the corresponding $R_\star$ conversion without
        this factor. The lower panel gives the absolute relative differences
        with respect to $\beta_\Gamma/H_p$.
    }
    \label{fig:beta-scans}
\end{figure}

As discussed in \cref{subsec:transition-parameters}, we use the mean bubble separation obtained from the integrated nucleation history, \cref{eq:bubble-density-Rstar}, and convert it to an inverse duration using \cref{eq:beta-Rstar-mapping}. \Cref{fig:beta-scans} compares this prescription with the local rate derivative in \cref{eq:beta-rate-definition} and with the same $R_\star$ conversion without the finite converted fraction.

The distinction becomes important for slow and strongly supercooled
transitions. The local derivative assumes that the nucleation rate can be
characterised by an approximately exponential growth around percolation.
When this approximation fails, $\beta_\Gamma/H_p$ no longer provides a
reliable measure of the characteristic bubble separation and can even
become negative if the nucleation rate is decreasing at $T_p$.
This behaviour, together with the relation between the local and $R_\star$-based definitions, is discussed in detail in \cite{Matuszak:2026xsz}.
The mean bubble separation instead remains well defined for the full nucleation history.
We therefore use $(\beta/H_p)_{R_\star,f_{\rm perc}}$ throughout our gravitational-wave analysis, including the finite converted-fraction correction appropriate to our percolation criterion.

\section{Numerical scan and PTA analysis}
\label{app:Scan-GWanalysis}

\subsection{Numerical scan and the input scale}
\label{app:numerical-rescaling}
\label{app:CW-scaling}
\label{app:percolation-scan}

We define the model parameters at the input scale $\mu_0$ as in \cref{eq:boundary-conditions}.
Varying $\mu_0$ keeping $g(\mu_0)$ fixed therefore changes the physical symmetry-breaking scale.
This should be distinguished from varying the thermal matching scales to estimate the perturbative uncertainty for a fixed model.
Using the minimum in \cref{eq:CW-vev}, the masses in the broken vacuum
are
\begin{equation}
    m_{A'}=gv=\ee^{1/6}\mu_0,
    \label{eq:CW-dark-photon-mass}
\end{equation}
and
\begin{equation}
    m_{h_D}^2
    \equiv
    \left.
    \frac{\mathrm d^2 V_{\rm CW}}{\mathrm d\phi^2}
    \right|_{\phi=v}
    =\frac{3g^4v^2}{8\pi^2}
    =\frac{3g^2}{8\pi^2}m_{A'}^2.
    \label{eq:CW-dark-Higgs-mass}
\end{equation}
Here $m_{A'}$ is the dark-photon mass and $m_{h_D}$ is the loop suppressed mass of the radial scalar, or dark Higgs. 

We use these zero-temperature masses to calculate the dark-sector contributions to $g_{*\rho}(T)$ and $g_{*s}(T)$, including the suppression of particles as they become non-relativistic.
At fixed $(g,\mu_0)$,the $\mathrm{NLO}_{\rm corr}$ and OPA scans have identical vacuum masses and $\Delta V_{\rm CW}$, given in \cref{eq:CW-deltaV}, because both use the same zero-temperature Coleman--Weinberg potential.

Using the scaling described in \cref{subsec:scale-scan}, we compute $S_{\rm grad,NLO}$ and $S_{Z_{\rm sp}}$ at $\mu_0=1\,\GeV$ for each $g$ and construct the corrected action using
\cref{eq:corrected-gradient-action}.
Since the correction depends only on $g$ and $x=T/\mu_0$, the resulting action table can be reused throughout the $\mu_0$ scan.
In the three-dimensional calculation, we use the dimensionless variables $\phi_3/\sqrt{\mu_0}$ and $\mu_0 r$, with the potential expressed in units of $\mu_0^3$.
The action contributions are tabulated at 256 logarithmically spaced values of $x$, with the low-$g$ tables extending down to $x=5\times10^{-12}$.

The requested parameter grid contains 75 values of $g$ and 102 values
of $\mu_0$ within \cref{eq:scan-range}. We use $\Delta g=0.005$ for
$0.4\leq g\leq0.55$, together with 40 uniformly spaced points between
$g=0.5625$ and $0.95$ and the exact benchmark couplings used in the
figures. The scale grid consists of 101 logarithmically spaced values,
supplemented by $\mu_0=1\,\GeV$. The $\mathrm{NLO}_{\rm corr}$ and OPA scans use the same parameter grid.

For OPA, we use the semi-analytic action of \cite{Levi:2022bzt}, with the running couplings and mass parameters evaluated at $\mu=\pi T$.
The OPA action is sampled at 1000
logarithmically spaced temperatures.
The $\mathrm{NLO}_{\rm corr}$ and OPA action tables are passed to the same rate construction, with $\Gamma=T^4e^{-S}$.

For each $(g,\mu_0)$, we determine the percolation temperature from $I(T_p)=-\log(0.71)$ using \cref{eq:false-vacuum-probability}.
We restrict the calculation to the temperature interval covered by the action table, without extrapolating the action.

We estimate the relative integration error in $R_\star$ by comparing two grid resolutions and flag points for which it exceeds $0.05\%$.
We also check that the integrand in \cref{eq:bubble-density-Rstar} is suppressed at the upper temperature limit.
For the $\mathrm{NLO}_{\rm corr}$ scan, we evaluate the completion diagnostic
$\mathcal C_p$ defined in \cref{eq:completion-diagnostic}.

We use the same prescriptions for the Hubble rate, percolation,
$R_\star$, and reheating in the $\mathrm{NLO}_{\rm corr}$ and OPA scans.
The GW spectra are computed using the central prescription of
\cref{subsec:GW-spectrum} and analysed with the same statistical method.
For the wall surface tension, we use the OPA polynomial in the OPA scan and the one-loop three-dimensional potential $\widehat V_3^{\mathrm{1L}}$ given in \cref{eq:app-one-loop-potential} for $\mathrm{NLO}_{\rm corr}$.
The different potentials therefore affect both the tunnelling action and the wall surface tension.
The determinant-calibrated correction itself modifies only the nucleation-rate exponent and does not change the wall tension.

\subsection{PTA likelihood and statistical inference}
\label{app:pta-likelihood}

A free-spectrum analysis assigns a separate amplitude to the signal common
to the pulsars at each sampled Fourier frequency, rather than assuming a
fixed spectral shape \cite{NANOGrav:2023hvm,Antoniadis:2022pcn}.
The released data products provide the corresponding probability
distributions as kernel-density estimates (KDEs) of the logarithmic
timing-residual amplitude, with the pulsar timing and noise uncertainties
already incorporated.
We use the corrected NG15 Hellings--Downs release v2.0.0 and the IPTA DR2
free-spectrum product, with the frequency selections specified in
\cref{sec:phenomenology}.
Our predicted gravitational-wave spectra are evaluated against these
distributions using \textsc{Ceffyl}~\cite{Lamb:2023jls}.

For comparison with the free-spectrum likelihood, each predicted spectrum
is interpolated onto the NG15 or IPTA DR2 frequencies, linearly in
$(\log f,\log[h^2\Omega_{\rm GW}])$.
We then convert it to the timing-residual power used by \textsc{Ceffyl} \cite{Lamb:2023jls},
\begin{equation}
\rho_k^2(\theta)
=
\frac{
H_{100}^2\,h^2\Omega_{\rm GW}(f_k;\theta)
}{
8\pi^4 f_k^5 T_{\rm span}
},
\qquad
H_{100}=100\,\mathrm{km\,s^{-1}\,Mpc^{-1}}.
\label{eq:ceffyl-rho}
\end{equation}
Here $\rho_k$ is the timing-residual amplitude in the frequency bin $f_k$, $\theta=(g,\mu_0)$ is the vector of model parameters and
$T_{\rm span}=1/f_1$, where $f_1$ is the lowest Fourier frequency.
The likelihood is evaluated using the predicted normalization of the
spectrum, without fitting an additional amplitude.

Writing $y_k=\log_{10}[\rho_k/\mathrm{s}]$, we denote the KDE in the $k$th frequency bin by $q_k(y_k)$.
Since the free-spectrum prior is uniform in $y_k$, these densities can be used directly in the refitting likelihood. Treating the frequency bins as independent gives
\cite{Lamb:2023jls}
\begin{equation}
\ln\mathcal L(d|\theta)
\simeq
\sum_{k=1}^{N_f}
\ln q_k\!\left[y_k(\theta)\right]
+\mathrm{const}.
\label{eq:ceffyl-free-spectrum-likelihood}
\end{equation}
The independence assumption applies only between frequency bins. 
The assumptions
about correlations among pulsars remain those of the underlying NG15 and
IPTA DR2 analyses.

We evaluate the tabulated KDEs by interpolating $\ln q_k$ linearly in
$y_k$.
Below the tabulated range, we keep the density fixed at its boundary value, following \cite{Lamb:2023jls}, while predictions above the upper boundary are assigned zero likelihood.
None of the NG15 or IPTA DR2 best-fit spectra for $\mathrm{NLO}_{\rm corr}$ or OPA reaches either boundary.
Points requiring the lower-boundary continuation account for at most $1.6\%$ of the posterior probability.

We evaluate the posterior directly on the grid of predicted spectra. Defining
$u=\log_{10}(\mu_0/\GeV)$, we take flat priors in $g$ and $u$, i.e. 
\begin{equation}
\pi(g,u)=\mathrm{const},
\qquad
\pi(g,\mu_0)\propto\frac{1}{\mu_0}.
\label{eq:pta-prior}
\end{equation}
The posterior probability associated with each grid point is
\begin{equation}
P_{ij}
=
\frac{
e^{\ell_{ij}-\ell_{\max}}\,
\Delta g_i\Delta u_j
}{
\displaystyle\sum_{m,n}
e^{\ell_{mn}-\ell_{\max}}\,
\Delta g_m\Delta u_n
},
\qquad
\ell_{ij}=\ln\mathcal L(d|g_i,\mu_{0,j}).
\label{eq:ceffyl-posterior}
\end{equation}
The cell widths are defined by the midpoints between neighbouring scan
values, with half-spacing extensions at the outer boundaries. The factors
$\Delta g_i\Delta u_j$ account for the non-uniform sampling in $g$ and
ensure that the additional point at $\mu_0=1~\GeV$ does not receive extra
prior weight. The normalization is performed separately for
$\mathrm{NLO}_{\rm corr}$ and OPA over the valid templates of each scan.
Missing or failed points are assigned zero weight and are not filled by
interpolation.

The best-fit point is the evaluated template with the largest
$\ell_{ij}$.
The one-dimensional posterior densities are obtained by
summing $P_{ij}$ over the other parameter and dividing by the corresponding
cell width.
The $68\%$ and $95\%$ credible regions are obtained by ranking the grid cells according to
\begin{equation}
p_{ij}
=
\frac{P_{ij}}{\Delta g_i\Delta u_j}
\end{equation}
and retaining those that together contain $68\%$ or $95\%$ of the
posterior probability.
The quoted best-fit coordinates therefore
correspond to points on the theory grid.

For each PTA data set, $\ln\mathcal L_{\max}$ is used to identify the
best-fit point within each scan.
We do not use differences in $\ln\mathcal L_{\max}$ to claim a statistical preference between $\mathrm{NLO}_{\rm corr}$ and OPA.
Points with $\mathcal C_p>0$ are retained in the analysis.
For $\mathrm{NLO}_{\rm corr}$, they account for approximately $9.2\%$ of the NG15 posterior probability and $0.7\%$ of the IPTA DR2 posterior probability, while neither best-fit point lies in this region.

\bibliographystyle{JHEP}
\bibliography{lib_jhep}

\end{document}